\documentclass[12pt,preprint]{aastex}
\slugcomment{To appear in the Astrophysical Journal}

\def\kmsec{\mbox{km~s$^{\rm -1}$}}
\def\teff{\mbox{$T_{\rm eff}$}}
\def\logg{\mbox{log ${\it g}$}}

\def\BmV0{\mbox{(B-V)$^{\rm o}$}}
\def\VmK0{\mbox{(V-K)$^{\rm o}$}}
\def\MV0{\mbox{M$_{\rm V}^{\rm o}$}}
\def\MV{\mbox{M$_{\rm V}$}}

\def\etal{\mbox{{\it et al.}}}
\def\eg{\mbox{{\it e.g.}}}
\def\ie{\mbox{{\it i.e.}}}

\shorttitle{Three-Dimensional Spectral Classification}
\shortauthors{Snider et al.}

\begin{document}

\title{Three-Dimensional Spectral Classification of Low-Metallicity
Stars Using Artificial Neural Networks}

\author{
Shawn Snider\altaffilmark{1},
Carlos Allende Prieto\altaffilmark{1},
Ted von Hippel\altaffilmark{2},
Timothy C. Beers\altaffilmark{3},
Christopher Sneden\altaffilmark{1},
Yuan Qu\altaffilmark{1},
Silvia Rossi\altaffilmark{4},
}

\altaffiltext{1}{Department of Astronomy and McDonald Observatory,
University of Texas, Austin, TX 78712; 
shawn,callende,yqu@astro.as.utexas.edu, chris@verdi.as.utexas.edu}

\altaffiltext{2}{Gemini Observatory, 670 North A`ohoku Place, Hilo, HI 96720;
ted@gemini.edu}

\altaffiltext{3}{Department of Physics and Astronomy, Michigan State 
University, East Lansing, MI 48824; beers@pa.msu.edu}
 
\altaffiltext{4}{Instituto Astron\^omico e Geof\'isico, Universidade de
S\~ao Paulo, Av. Miguel Stefano 4200, 04301-904, S\~ao Paulo, Brazil;
rossi@orion.iagusp.usp.br}

\begin{abstract}

We explore the application of artificial neural networks (ANNs) for the
estimation of atmospheric parameters (\teff , \logg , and [Fe/H]) for Galactic
F- and G-type stars.
The ANNs are fed with medium-resolution 
($\Delta \lambda \sim 1-2$ \AA\ ) non flux-calibrated 
spectroscopic observations.
From a sample of 279 stars with previous
high-resolution determinations of metallicity, and a set of (external)
estimates of temperature and surface gravity, our ANNs are able to predict
\teff\ with an accuracy of $\sigma (\teff) = 135-150$ K over the range $4250 \le
\teff \le 6500$ K, \logg\ with an accuracy of $\sigma (\logg) = 0.25-0.30$ dex
over the range $1.0 \le  \logg\ \le 5.0$ dex, and [Fe/H] with an accuracy
$\sigma ({\rm [Fe/H]}) = 0.15-0.20$ dex over the range $-4.0 \le {\rm [Fe/H]} \le
+0.3$.  Such accuracies are competitive with the results obtained by fine
analysis of high-resolution spectra.  It is noteworthy that the ANNs are able
to obtain these results {\it without} consideration of photometric information
for these stars.  We have also explored the impact of the signal-to-noise ratio
(S/N) on the behavior of ANNs, and conclude that, when analyzed with ANNs
trained on spectra of commensurate S/N, it is possible to extract physical
parameter estimates of similar accuracy with stellar spectra having S/N  as low as 13.  Taken together, these results indicate that the ANN approach should
be of primary importance for use in present and future large-scale
spectroscopic surveys.

\end{abstract}

\section{Introduction}

Many important problems in Galactic and extragalactic astronomy can only be
constrained through the acquisition of extremely large databases of low- and/or
medium-resolution spectroscopy.  Efficient multi-object spectrometers are now
in routine operation, \eg, WIYN's and CTIO's Hydra (Barden \etal\ 1993), AAT's
2dF (Gray \etal\ 1993), Lick Observatory's AMOS (Brodie \& Epps 1993), WHT's
WYFFOS (Bingham \etal\ 1994), and the 6DF on the UK Schmidt telescope (Watson,
Parker, \& Miziarksi 1998).  Spectrographs capable of obtaining several hundred
spectra at a time are now reality, such as that used in the Sloan Digital 
Sky Survey (SDSS; York \etal\ 2000), with others planned for installation 
at many telescopes in the
near future.  These instruments can rapidly assemble libraries of
10$^3$--10$^5$ spectra even during the course of a single night or single
observing run; new spectrographs with even greater multiplexing advantages are
in various stages of development.  Although aimed at the study of galaxies and
quasars, the on-going SDSS  will amass 
about  $10^4$ stellar spectra  with a resolving power of $R=2000$  between 
3900 and  9100 \AA\ in a field around the North Galactic Pole. 
The combination of  micro-arcsecond-accuracy
astrometry with $R \simeq 3700$ spectroscopy for
$\sim 10^8$ stars that will be available in the future from GAIA (see
Perryman \etal\ 2001)  will revolutionize our understanding of the
dynamics and the chemical evolution of the Milky Way and neighboring
galaxies.

The extraction of useful physical information from these large spectral
databases (for stars, parameters such as effective temperatures, surface
gravities, metallicities, elemental abundance ratios, and radial velocities)
can of course be done one-star-at-a-time using well-understood analysis
techniques --- but this requires a small army of researchers.  A much more
sensible approach is to adapt and develop new techniques for automatic,
accurate, and efficient extraction of key physical information from the
spectra, ideally in real time.

A number of previous authors have pursued the development of methods for
obtaining estimates of atmospheric parameters from low- to medium-resolution
stellar spectra.  Jones (1966), for example, in early pioneering work, made
visual estimates of ten line ratios and six line strengths for a uniform set of
photographic Coud\'e spectra obtained with the Palomar 200" telescope.  He
then performed a principal component analysis of these data, and calibrated the
three largest principal components with temperature, luminosity, and metal
abundance.  Th\'evenin \& Foy (1983) explored a technique based on the
comparison of measured equivalent widths for several prominent spectral lines
in $\Delta \lambda \sim 2$ \AA\ resolution spectra with grids of
theoretical equivalent widths
obtained from model atmospheres.  Although their test sample of stars was
rather small, the resulting derived errors were certainly respectable.  Cayrel
\etal\ (1991) pursued similar ideas, making use of a matching algorithm that
compared relatively high  S/N (40 to 100), low-resolution ($\Delta
\lambda \sim 5$  \AA\ ) spectra with grids of synthetic spectra, and again
achieved encouraging results based on a small number of comparison stars
covering a wide range of atmospheric parameters.  Cuisinier \etal\ (1994)
 and Gray, Graham \& Hoyt (2001) described several techniques for 
 the estimation of stellar parameters from medium-resolution spectra, 
 based on comparisons with a grid of model atmospheres, but their 
 application was mostly to more metal-rich stars of the Galactic disk 
 populations.

In this paper we examine the merits of a particular type of expert system based
on back-propagation artificial neural networks (hereafter ANNs) for
astrophysical parameter estimation from medium-resolution stellar
spectroscopy.  Previously explored parameter-estimation techniques
include cross-correlation and maximum-likelihood fitting (Katz \etal\ 1998) and
minimum vector-distance estimation (Kurtz 1984).  Line-fitting techniques
depend on prior knowledge of approximate spectral types before the
determination of which lines to fit can be made, since different species can
absorb at the same wavelengths in different (\teff, \logg) domains.
Cross-correlation, in its simplest form, also weights comparisons by line
strength, although the strongest lines are not necessarily the features with
the highest weight in classification assignment. Cross-correlation also requires a well-populated library
of homogeneous quality.  Minimum vector-distance techniques have had some
success, but have not been pursued to the level desired for our purposes.  It
is important to note that these classical techniques are based on the
application of {\it linear} operations.  Since we expect to find rather subtle
{\it non-linear} relationships between temperature, surface gravity,
and metallicity indicators in a given stellar spectrum, classification schemes
that allow non-linear relationships between parameters, such as ANNs, should
offer significant advantages. 

Supervised ANNs have application to a wide variety of non-linear optimization
problems.  For the estimation of stellar atmospheric parameters, a growing body of
work (e.g., Gulati et al. 1994; von Hippel et al. 1994; Vieira \& Ponz 1995;
Weaver \& Torres-Dodgen 1995, 1997; Bailer-Jones et al. 1997, 1998) has
demonstrated that automated ANNs can be robust and precise classifiers of
stellar spectra.  Recently, Bailer-Jones (2000) has explored the capability of
ANN techniques to deduce \teff, \logg, and [Fe/H] for stars to be observed with
the medium- and broad-band photometric systems to be implemented for the GAIA
space mission.  Rhee, Beers, \& Irwin (1999) and Rhee (2000) discussed the
development and implementation of an ANN approach for the analysis of digital
scans of the HK-survey objective-prism plates, and found that, with the
addition of rough color information from calibrated photographic surveys, they
were able to select metal-deficient stars without the introduction of
bias related to the temperature that plagued the original visual-selection
technique.  Allende Prieto \etal\ (2000) applied an ANN approach to sets of
prominent line indices in medium-resolution spectra from the HK survey, and
demonstrated that reasonably accurate estimates of [Fe/H] and broad-band color $(B-V)_0$  could be obtained in this way.  We refer interested readers to these
papers, and references therein, for both general information on ANNs and for
specific mathematical details of their training and testing.

In this paper we demonstrate the utility of the ANN approach for the analysis
of medium-resolution spectra of metal-poor stars of the Galactic halo and
thick-disk populations.  Most of the previous automated stellar spectral
classification efforts have focused on local samples of stars with
metallicities characteristic of the Galactic disk.  We note, however, that
Prugniel \& Soubiran (2001) have recently provided a large database of high-
and low-resolution spectra (including stars with metallicities as low as [Fe/H]
= --2.7) obtained with the ELODIE spectrograph on the OHP 1.5m telescope, and
are clearly in the process of further developing the TGMET spectral
parameterization method of Soubiran \etal\ (1998) and Katz \etal\ (1998).  

In \S2 the dataset for training and testing the ANNs is described, and in \S3
the preparation of the database for ANN input is outlined.  The assignment of
``known'' atmospheric parameters for the stars in our sample is discussed in \S4.
A detailed description of our adopted ANN methodology, and the results of its
application to stellar spectra, are provided in \S5. In this same section we
explore the impact of spectral S/N on the derivation of atmospheric parameters
through a series of numerical experiments.  Our conclusions and suggestions for
future work are provided in \S6.  In the Appendix we discuss the small number
of deviant cases that were noted during the course of our analysis.

\section{The Spectroscopic Database}

The stars that form the basis of our evaluation of the ANN approach
were observed during medium-resolution spectroscopic campaigns by Beers
and collaborators for the metal-poor stars of the HK survey (Beers
\etal\ 1985, 1992).  A discussion of the various campaigns is given in
Beers (1999); in Table~1  we list the parameters of the
spectra employed in the present study.  To improve the homogeneity, we
limited our sample to the best 6 telescope-detector combinations, from
the 12 considered by Beers \etal\ 1999.  This filtering reduces the
number of standards from the more than 500 studied by Beers \etal\ to
279 stars.  Columns (1) and (2) of the table list the telescope and
detector used.  Column (3) lists the wavelength coverage of the spectra
obtained.  Column (4) lists the dispersion of the spectra, in some
cases after a re-binning was employed during the initial data
reduction.  Column (5) lists the total number of spectra contributed to
this study for each of the various combinations.

The stars that comprise our study are a subset of the calibration stars used in
the Beers \etal\ medium-resolution surveys (see Beers \etal\ 1999).  They were
selected to cover the range of metallicities, temperatures, and surface
gravities (see \S4) expected to pertain to the metal-poor stars discovered in
the extensive HK survey.  Thus, these were the template stars used to judge the
accuracy of the atmospheric parameters (in particular the metallicity) derived for
candidate HK survey low-metallicity stars.  Hence, all our program stars  have
available estimates of [Fe/H]. We employ the standard abundance
notation that [A/B]~$\equiv$~log$_{\rm 10}$(N$_{\rm A}$/N$_{\rm B}$)$_{\rm
star}$~-- log$_{\rm10}$(N$_{\rm A}$/N$_{\rm B}$)$_{\odot}$, and equate
metallicity to the stellar [Fe/H] value from previous analyses of
high-resolution spectroscopy by many workers.  We have supplemented this
information with newly derived estimates of \teff\ and \logg\ from several
techniques, as described below.  

Beers \etal\ (1999) describe a method for the estimation of stellar
metallicity from medium-resolution ($\Delta \lambda \sim $ 1--2~\AA\ ) spectroscopy
and broad-band $(B-V)_0$ colors. This technique makes use of empirical
corrections, based on standard stars of known abundance, to the predicted line
strengths from synthetic spectra and estimated broad-band colors from model
atmospheres.  The final calibration obtained by Beers \etal\ (1999) provides
the means for accurate estimation of metallicity ($\sigma$([Fe/H])~$\sim$
0.15--0.20~dex) over the entire range of metallicities of known Galactic stars
(--4.0~$\leq$~[Fe/H]~$\leq$~+0.3).  This represents a clear improvement over
the Beers \etal\ (1990) calibration, which had difficulty in obtaining
metallicity estimates for stars with [Fe/H]~$\geq$ --1.0 due to saturation of
the \ion{Ca}{2} K-line they used as their primary abundance indicator.
However, there still are limitations to the Beers \etal\ (1999) approach.   For
instance, the use of multiple levels of empirical corrections makes the
approach somewhat cumbersome to implement for general use.  This is one of the
reasons we have begun to explore the use of ANNs for future work.  Although
both the ANN approach and the Beers \etal\ method are capable of providing
accurate metallicity estimates, we demonstrate below that the ANN approach can
obtain a similar level of accuracy using non flux-calibrated spectra {\it
without the need for additional broad-band photometric observations}, and it is
largely insensitive to reddening.
Furthermore, our ANN technique is also capable of estimating temperatures and
surface gravities, which the Beers \etal\ calibration did not provide.

\begin{figure}
\epsscale{0.8}
\plotone{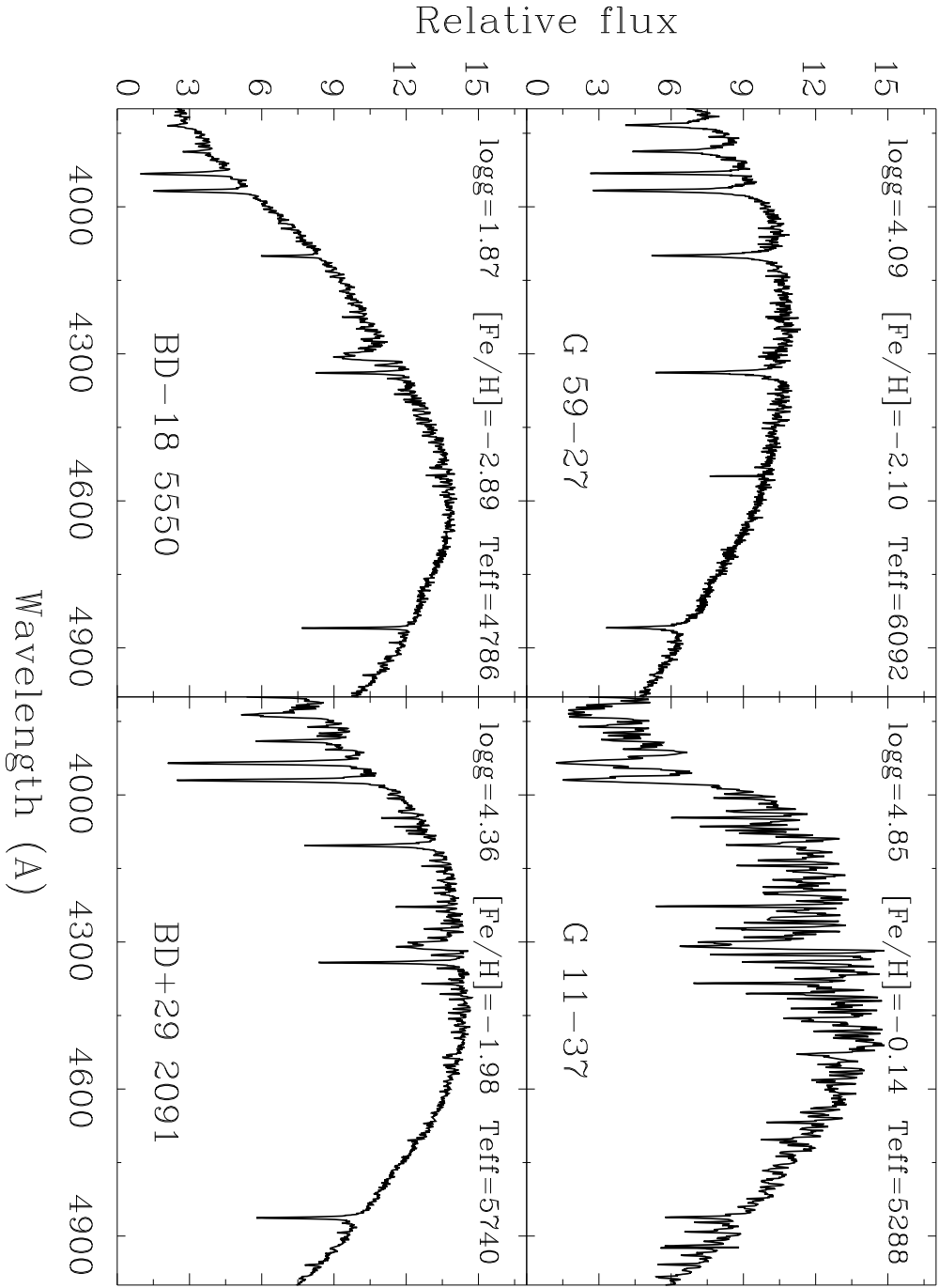}
\figcaption{
Examples of original {\it raw} spectra for four program stars covering a variety
of \teff, \logg, and [Fe/H]. 
\label{fig-rawspectra}}
\end{figure}

The reduction and analysis of the spectroscopic data are described in Beers
\etal\ (1999), and will not be repeated here.  Because this is our first
attempt at developing a medium-resolution neural network for future use, we
decided to impose a rather severe lower limit on the S/N ratios of the
spectroscopic data that were used for the construction of our training and
testing sets.  In order to be used in our analysis, a stellar spectrum was
required to have S/N~$>$~20 at 4000~\AA, and cover at least the wavelength
range 3850 to 4450~\AA.  As part of the selection process for program stars,
care was taken to make certain that any spurious features, such as cosmic ray
hits, were removed from each spectrum prior to assembly of our data sets, since
our ANN uses the entire spectrum in its analysis.  In the end, spectra of 279
stars were chosen for the ANN experiments described in this paper.  Several
examples of the raw (extracted and wavelength calibrated) spectra, prior to
their preparation for the ANNs, are shown in Figure~\ref{fig-rawspectra}.

\section{Unification of the Spectroscopic Data}

Successful application of the ANN techniques described below first requires the
creation of a dataset that is as uniform as possible.  For our purposes this
means manipulating the spectra until: (a) they are all on the same
stellar rest-wavelength scale, with identical starting and ending wavelengths;
(b) they have closely matched spectral resolutions; and (c) their observed
fluxes have been rectified in a consistent manner.  The steps taken to
transform the raw spectra into a form acceptable for ANN analysis are described
in this section.  For all of these steps we employed various tasks contained within
the IRAF\footnote{ IRAF is distributed by the National Optical Astronomy
Observatories, which are operated by the Association of Universities for
Research in Astronomy, Inc., under cooperative agreement with the National
Science Foundation.} software package.

First, the spectra were continuum flattened, effectively cancelling the
combination of the stellar spectral energy distribution (SED) and the
instrumental response.  Although the (potentially useful) stellar SED
is therefore destroyed, it would not have been possible to recover this
information for the majority of our stars due to the lack of available
flux calibrations for most of the spectroscopic observations obtained
during the HK survey campaigns.  The most important part of this step
was to treat the varied pseudo-continua of the raw spectra in a uniform
manner.  We were aided by the general weak-lined nature of our
metal-poor program spectra, which allowed reasonable identification of
regions that were relatively free of absorption features in wavelength
domains from the red to as blue as 3900~\AA.  At shorter wavelengths, a
conspiracy of increasing spectral line density, decreasing stellar
flux, and decreasing instrumental efficiency generally resulted in low
S/N and larger uncertainties in the placement of the pseudo-continuum
level.  We experimented extensively with different continuum
rectification techniques, and eventually found that  discarding the
strongest absorption features and repeatedly applying a smoothing
filter  worked best for our spectra as a whole.  This technique 
 did not produce satisfactory results for the few very
carbon-rich stars with strong bands of CH and C$_2$ (\eg, CS~22957-027:
Norris, Ryan, \& Beers 1997; Bonifacio \etal\ 1998) for which continuum
placement based on intermediate-resolution observations 
is difficult with any technique.
As a result, the carbon-enhanced stars were not used in the present
study.

The continuum-flattened spectra were then shifted to a common radial velocity.
Since uniformity in the velocity frame is crucial, but the zero-point is not,
all of our spectra were shifted in velocity to a single program-star template
spectrum, using the IRAF task  {\it dopcor}.  Cross-correlations of all spectra
with the template were done with the task {\it fxcor}.  The template spectrum
was chosen to be that of HD~122563 (F8IV), which contains features common to all of
the spectra being prepared. The program stars in our study
range from warm main-sequence dwarfs to cool red giants, and the stellar
metallicities have a  range of three orders of magnitude.  We note that 
for future work on
more extensive data sets, it may be best to use different velocity templates
for different \teff/\logg/[Fe/H] domains, and carefully tie the templates to a
common system.  We found this additional step unnecessary for the initial
exploration of ANNs considered here.

\begin{figure}
\epsscale{0.8}
\plotone{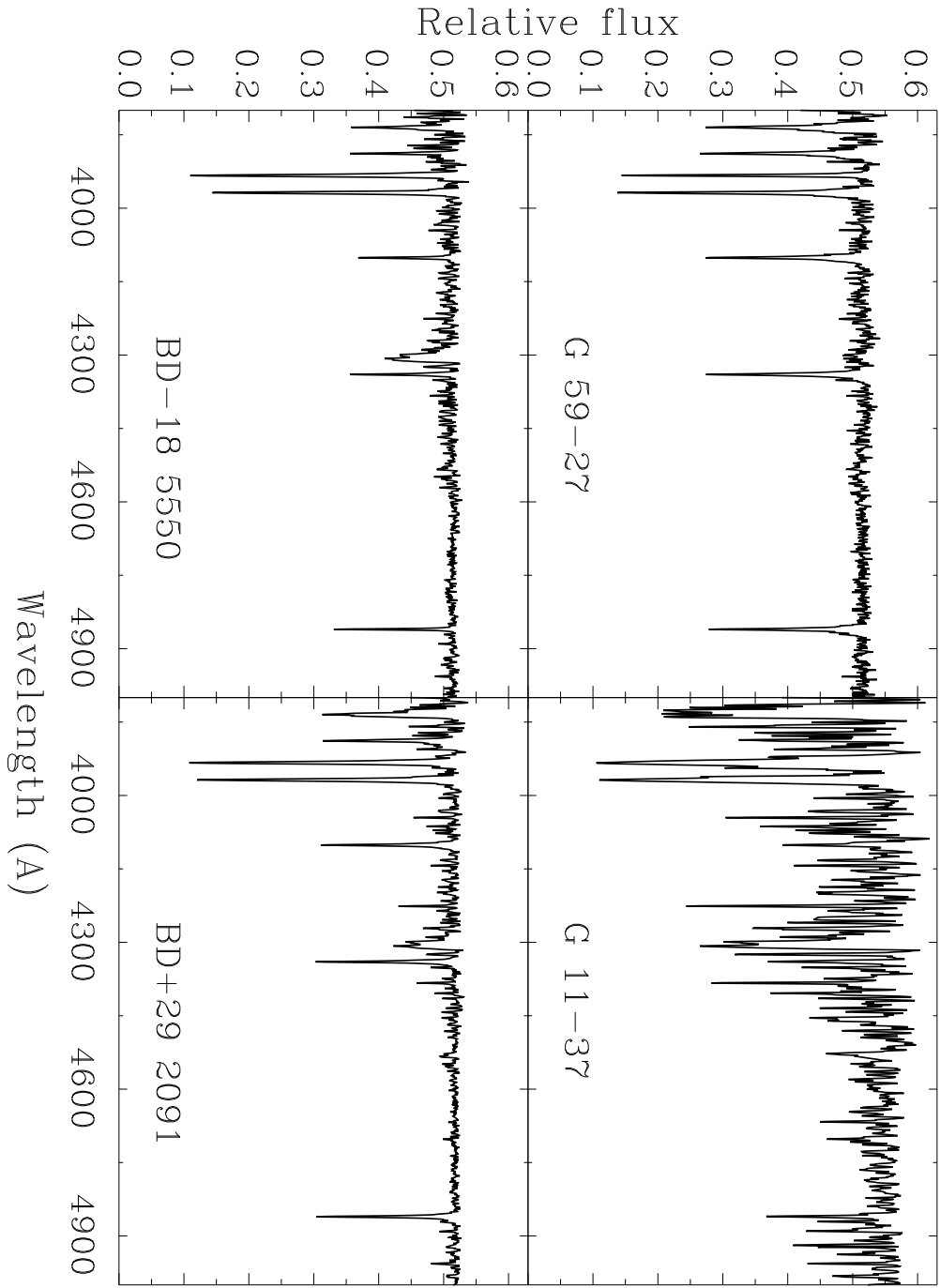}
\figcaption{
Examples of the same program stars as appeared in Figure 1, but after
preparation for the ANNs.  
\label{fig-finalspectra}} 
\end{figure}

Next, the spectra were re-binned to a common wavelength binning.  
We adopted a fixed
binning of 0.65~\AA/pix (the dispersion of the template star, and 
the majority of our spectra), and used the IRAF task {\it dispcor} to re-bin
all of the program stars.  The resolution is two to three times larger than
  the dispersion, depending on the data source and the particular 
  observing conditions.  With the same task,   we trimmed the spectra to a
common wavelength range, as our ANN can perform only on data sets of identical
wavelength coverage.  Using the IRAF task {\it wspectext}, we converted the
spectra to a text format acceptable to the ANN.  Finally, we multiplied the spectra by a constant factor to have an
average value of 0.5 in a selected wavelength range, since
 our ANNs were developed to be most sensitive to flux
values between 0 and 1.  In Figure~\ref{fig-finalspectra} we show the fully
modified spectra, ready for ANN input, of the same stars whose raw spectra
appear in Figure~\ref{fig-rawspectra}.

\section{Atmospheric Parameters of the Program Stars}

Beers \etal\ (1999) compiled and averaged metallicities that had been
determined from high-resolution spectroscopic analyses, based on flux-constant
plane-parallel LTE model atmospheres, for over 500 stars of the Galactic halo
and thick-disk populations.  We  selected our testing and training sets
from that pool, taking particular care to achieve a reasonably complete
distribution throughout the parameter space of \teff, \logg, and [Fe/H].
Although, in principle, an averaged set of \teff\ and \logg\ determinations
from the high-resolution analyses could have been used in our application, we
have chosen not to take this approach.  The \teff\ and \logg\ employed in the
high-resolution analyses come from a wide variety of sources (broad- and/or
narrow-band photometry, fits to isochrones, or fine analysis of the spectra
themselves), hence a more homogeneous set of temperatures and surface gravities
is desirable.  This decision also permits a comparison of our derived
parameters with independently obtained estimates from the high-resolution work.

Effective temperatures were derived from the $(B-V)$ colors, $E_{B-V}$
reddening corrections, and metallicities compiled by Beers \etal\ (1999).  For
dwarfs and subgiant stars, we  applied the calibrations of Alonso, Arribas, \& Mart\'{\i}nez-Roger (1996).  For more evolved giant stars, we  used the  Alonso, Arribas, \& Mart\'{\i}nez-Roger  (1999)
calibrations.  These calibrations are based on the Infrared Flux Method (IRFM),
developed by Blackwell and collaborators (see Blackwell \& Lynas-Gray 1994, and
references therein).  The IRFM compares the observed ratio of the bolometric
and monochromatic flux in the infrared with the ratio predicted by model
atmosphere analyses.  Since the effective temperature defines the bolometric
flux, the method only relies on the models to estimate the monochromatic flux
in the infrared.  From the use of several infrared photometric bands, it is
possible to check for internal consistency, which turns out to be exceptionally
good for \teff~$\gtrsim$~4500 K, resulting in mean errors  of only 1--2\%.  The
standard deviations of the polynomial fits to the IRFM \teff\ as functions of
$(B-V)_0$ color and metallicity  are in the range 100--170~K over the entire
parameter space relevant to our program stars.

For their main-sequence, subgiant, and giant stars, Beers \etal\ (1999) derived
absolute magnitudes and distances by making use of the revised Yale isochrones
(Green 1988; King, Demarque, \& Green 1988), over the metallicity interval
--3.0~$\leq$~[Fe/H]~$\leq$~0.0, and assuming ages between 5 and 15~Gyrs.  To
provide estimates for horizontal-branch stars and asymptotic giant-branch
stars, they adopted a relation between $M_V$, [Fe/H], and $B-V$ (see Beers
\etal\ for more details).  We  used their absolute magnitudes to
interpolate in the oxygen-enhanced  isochrones of Bergbusch \&
VandenBerg (1992), and derived bolometric corrections and stellar masses.  The
calculated luminosities were then combined with the effective temperatures from
the Alonso \etal\  calibrations to obtain stellar radii, and then with the
masses to derive estimates of the surface gravities.  This procedure involved
the adoption of an age to select the appropriate isochrone.  The age was set at
15~Gyr for the more metal-poor stars ([Fe/H]~$<$~--1.1), at 4~Gyr for those
with [Fe/H]~$>$~+0.03, and a linear variation between the extremes, fitting the
trend found by Edvardsson \etal\ (1993).

\begin{figure}
\epsscale{0.8}
\plotone{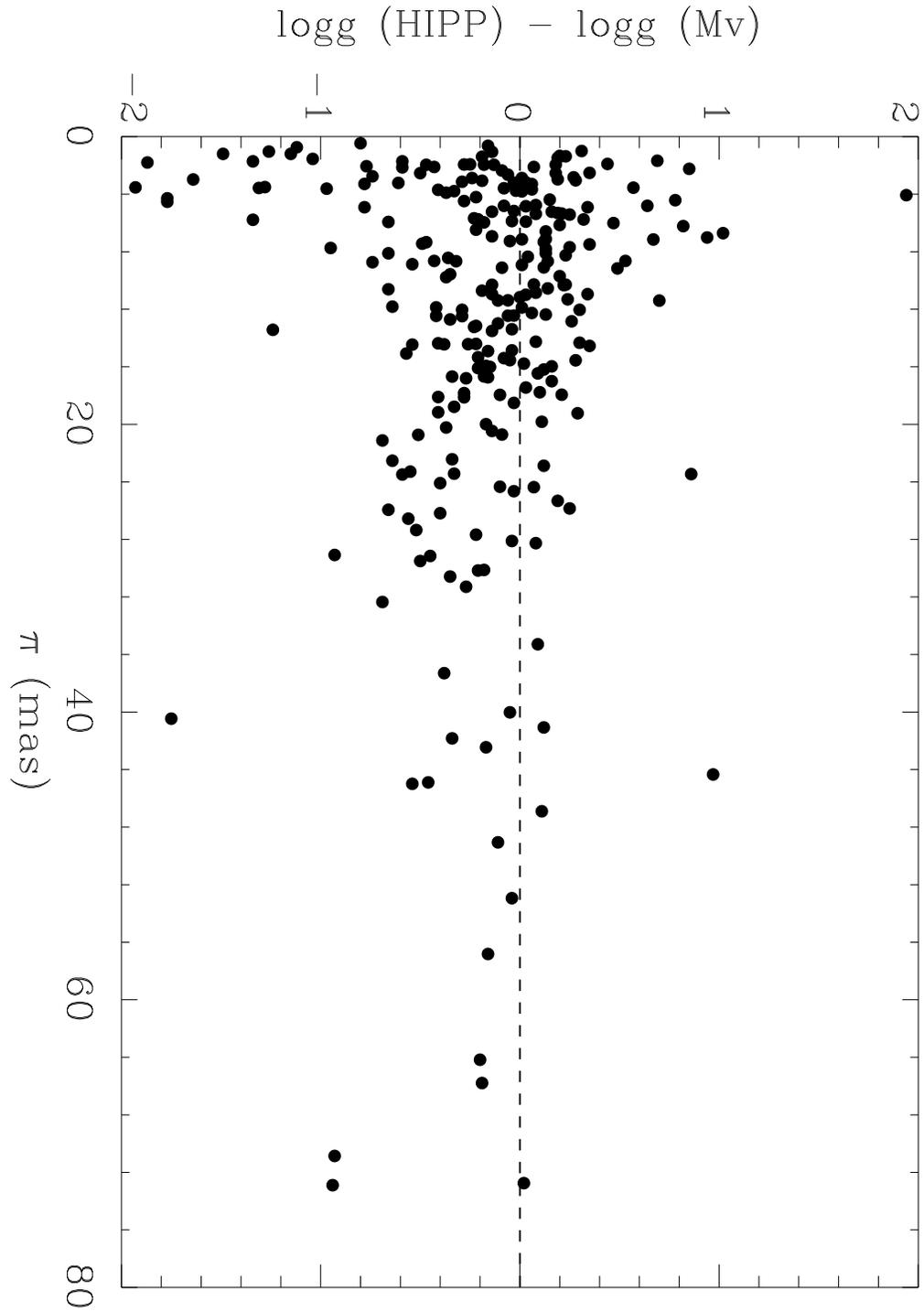}
\figcaption{A comparison of the differences in estimated surface gravity for
program stars based on values inferred from the {\it Hipparcos} distances, and
based on the $M_V$ reported by Beers et al. (1999), and described in
the text.   
\label{fig-gravtest}}
\end{figure}

Gravities can also be estimated by making use of the trigonometric parallaxes
($\pi$), in combination with the isochrones, as described in Allende Prieto
\etal\ (1999) or Allende Prieto \& Lambert (1999), although the relatively low
accuracy of present parallax measurements limits the validity of the
procedure to a small subset of our stars.  In Figure~\ref{fig-gravtest} we
illustrate the comparison between $\pi$-based and $M_V$-based gravity estimates
for the stars analyzed by Beers \etal\ (1999) with available {\it
Hipparcos} parallaxes (ESA 1997).  The errors in the parallaxes dominate
the discrepancies for stars farther away than 100~pc ($\pi$~$<$~10~mas).  For 115 (generally  metal-rich) dwarfs closer to the Sun than 100 pc, the mean
difference between the $\pi$-based gravities and the $M_V$-based gravities is
$<$\logg$_\pi$~--~\logg$_{MV}>$~= --0.15 ($\sigma $~=~0.31) dex.  The lack of
nearby evolved stars in our sample precludes the application of the same test
to them.

The atmospheric parameters \teff, \logg, and [Fe/H] adopted for each
star in our study will hereafter be referred to as ``catalog'' (CAT) values.
In Table~2  the catalog values for the set of stars used to
{\it train} the ANNs are listed as \teff\ $_{\rm CAT}$, \logg\ $_{\rm CAT}$,
and [Fe/H] $_{\rm CAT}$; the parameters used to test the ANNs are listed under
the same names in Table~3. Note that column (2) of each table
lists the source of the spectrum for each of our program stars, according to:
E = ESO 1.5m; K = KPNO 2.1m; L = LCO 2.5m; O =  ORM 2.5m; P = PAL 5m; S = SSO 2.3m\footnote{ESO $\equiv$ European Southern Observatory (Chile); KPNO $\equiv$ Kitt Peak National Observatory (USA); 
LCO $\equiv$ Las Campanas (Chile); ORM $\equiv$ Observatorio del Roque de los Muchachos (Spain); PAL $\equiv$ Mount Palomar (USA); SSO $\equiv$ Siding Spring Observatory (Australia) }.  
An asterisk next to the listed source indicates that the star is a
member of the nearby subsample described below.  A colon (:) next to the
catalog or network parameters indicates a large discrepancy; see the Appendix
for discussion of individual cases.

\section{Application of an Artificial Neural Network}

\subsection{Initiating and Running the ANN code}

In this work we have employed a back-propagation ANN code kindly made available
by B.D. Ripley (see Ripley 1993a, b).  Back-propagation is a standard ANN
training technique, though Ripley has implemented a few clever additions that
allow his code to operate without the free parameters of momentum and the
learning coefficient. ANN training is based on multi-dimensional
minimization techniques that converge to the desired solution by iteratively
providing the direction, but not the magnitude, of the updates necessary for
the many weights connecting the ANN nodes.  A learning coefficient is commonly
used to set the magnitude of the weight updates, whereas the momentum term sets
the degree to which the weight updates in the current iteration are related to
prior weight updates.  Careful tuning of the learning coefficient and the
momentum term are necessary in standard back-propagation codes when the
solution space has a number of local minima.  The only remaining free
parameters are the initial random weights interconnecting the various layers of
the ANN, and the criterion for stopping the learning process -- more discussion
of these is provided below.  Our network architectures were also standard and
fully connected from the input layer to one, and sometimes a second, hidden
layer and then to the output layer.  The connections between nodes are
numerical weights that contain the knowledge of the classification system.  The
training step involves adjusting the weights so that the ANNs provide a
generalized mapping of the input space (in our case, stellar spectra) to the
output space (in our case, the estimated atmospheric parameters of \teff, \logg,
or [Fe/H]).  The number of nodes in the input layer was dictated by the number
of spectral elements per spectrum.  In other words, our ANNs ingest
spectra, not derived parameters.  The number of output nodes was always one.

Experimentation demonstrated that ANNs designed to ambitiously fit two or
more desired parameters simultaneously (here, \teff, \logg, and [Fe/H]),
converged on a robust solution far less frequently than those that specialized
in a given parameter, \eg\ ,  \teff.  Thus, we built separate ANNs to determine
each of the atmospheric parameters.  Decisions on the appropriate number of hidden
nodes, and on whether to employ two layers of hidden nodes or just one, are
dictated by the level of complexity and non-linearity in mapping the input
space to the output space.  We experimented with a wide range of numbers of
hidden nodes in one layer (3, 5, 7, 9, 11, and 13) and in two layers (3:3, 5:5,
and 7:7).  We chose odd numbers of hidden nodes in order to span a wider range
of ANN complexity without having to train as many ANNs.  It is an important
general rule of ANN applications not to use too many hidden nodes, or the
number of free parameters grows too large and the ANNs just memorize their
training set rather then converge to the desired mapping.

As explained in \S2, the spectra in our present application were obtained at a
variety of observatories on a number of different spectrographs with a range of
wavelength coverages.  We wished to explore whether the heterogeneity in the
data sources would limit the quality of the classifications.  We thus chose to
train ANNs on the entire training set (hereafter, the total/full sample), as
well as three additional subsamples.  One subsample (nearby/full) was comprised
of 101 stars with parallaxes larger than 10 mas, as measured by {\it Hipparcos},
 and therefore with a fairly well constrained M$_V$ (see \S4).  
Another subsample (total/kpno) consisted only of
data obtained at the KPNO 2.1m telescope, since this was the most common source
for our spectroscopy.  The final subsample (nearby/kpno) was the intersection
between the nearby and the KPNO subsamples.  For the KPNO data the available
wavelength range was 3733.9 to 4964.5~\AA, corresponding to $1906$ input
parameters; for the other two subsamples the available wavelength range was
3836.6 to 4452.9~\AA, corresponding to $955$ input parameters.  
Table~4 lists the number of stars contained in each subsample.
Below we point out that, in some instances, the additional information provided
by the extended spectral coverage in the KPNO spectra results in superior
performance of the ANNs, which is perhaps no surprise.

Since our data sets were small, we used approximately 75\%  of the data,
randomly drawn, to train the ANNs and the remaining 25\%  of the data to test
the ANNs.  Standard practice is to divide the data set in half for training and
testing, but other divisions are acceptable as long as the training and testing
sets span the same regions of classification space.  As shown in
Figure~\ref{fig-paramspace}, our training and testing data essentially satisfy
these criteria.  A few spectra near the limits of the parameter domains will
force the ANNs to extrapolate, but objects with the most extreme parameters
will be difficult to classify by any automated technique.

We found little difference between the results of most of the ANNs that we
tried, indicating that this classification problem is well posed, has a broad
global minimum (optimal solution), and the data are appropriate for the task.
Rather than present the results for all ANNs that we built, we will concentrate
on the architectures with one hidden layer and 9 hidden nodes for \teff\ and
\logg\ and 13 hidden nodes for [Fe/H].  These particular architectures yielded 
the best classifications.  For
comparison, the architectures which led to the worst classifications still
provided adequate results, however, with errors larger by only 50 K (\teff),
0.04 dex (\logg), and 0.15 dex ([Fe/H]).

The input-to-output mapping function that is used in the nodes saturates when
the total input to a given node (the sum of the individual inputs multiplied by
their weights), approaches $0$, or becomes significantly greater than $1$.  For
this reason the initial range of the weights connecting the nodes is from $-1$
to $+1$, and the actual range of atmospheric parameters need to be remapped to
values between $0$ and $1$.  Given the ranges for our program stars in
effective temperature (4000~K~$\lesssim$~\teff~$\lesssim$~6500~K), surface
gravity (0.0~$\lesssim$~\logg~$\lesssim$~6.0), and metallicity
(--4.0~$\lesssim$~[Fe/H]~$\lesssim$~+0.5), we re-mapped these parameters with
the following simple equations:

   \teff$_{\rm re-map}$ = (\teff\ -- 4000) / 2500

   (\logg$)_{\rm re-map}$ = (\logg)\ / 6

   [Fe/H]$_{\rm re-map}$ = ([Fe/H] -- 0.5) / --4.5.

We arbitrarily set the maximum number of learning iterations to 1000, stopping
the ANN training at that point.  Since the major portion of the error
minimization occurs in the first few dozen iterations, followed by an
exponential decrease in the rate of learning, by the time the networks had
trained to 1000 iterations the learning rate was essentially zero.  Although
the training of a given ANN architecture with 1000 iterations took only 30
minutes on a {\it Sun Ultra 30} workstation, the exploration of a range of
architectures (and details of the data set) required weeks of computer time.
Note, however, that classifying data with a trained ANN is much faster than
training the ANN; an individual classification requires less than one
second of CPU time.

We explored ten different initial random-weight configurations for each of the
final ANNs we chose to apply.  By training the same architecture with different
initial random weights we obtained a measure of how likely the ANNs were to
converge on local minima rather than the desired global minimum.  The \teff\
ANNs were the easiest to train, and converged on spurious local minima in only
one out of ten instances.  The \logg\ problem was more challenging, especially
when training on the smaller nearby sample.  For this parameter the ANNs
converged on local minima in seven out of ten instances.  Since
cross-validation, {\it i.e.} testing with the unseen data set, makes it readily
apparent when the ANNs converge on local minima, it is easy to retain and apply
only the well-trained ANNs.  The difficulty that our ANNs experienced when
training on the small data sets indicates that we are approaching the lower
limit on the appropriate number of spectra for the training step.  We also
expect the data heterogeneity to be a factor in making it more difficult for
the ANNs to train.  While ANNs have the advantage over many other techniques in
that they can be trained to ignore data heterogeneity, the training procedure
is certain to be improved by the provision of more examples, and
thus avoiding unwanted correlations between the input and output spaces.

\subsection{Results}

In Tables~2  (the training set) and 3 (the
testing set) the atmospheric parameters computed by our best ANNs are listed as
\teff\ $_{\rm ANN}$, \logg\ $_{\rm ANN}$, and [Fe/H]$_{\rm ANN}$, respectively.
Figure~\ref{fig-teffann} presents the results for our \teff\ ANNs for the four
subsamples.  The points plotted as asterisks represent the training set;
open circles the testing set.  Naturally, the training set displays a better
distribution about the correspondence line, exhibiting both a lower mean
residual and a lower scatter.  The statistics of the residuals are discussed
below.  Note that there are a few deviant \teff\ classifications, especially in
the total/full subsample, the most heterogeneous of the data sets.  The stars
with the most deviant classifications are discussed in further detail in the
Appendix.

Figure~\ref{fig-loggann} presents the results for our \logg\ ANNs.  Note that
the axes of panels (a) and (b) display a much smaller range of \logg\ values
than the axes of panels (c) and (d), since the nearby star sample does not
include any giants.  Since the \logg\ classifications are based on a few 
spectral features, the level of scatter seen in the figures is to be expected.
As before, the total/full subsample exhibits the most deviant classifications.

Figure~\ref{fig-fehann} presents the results for our [Fe/H] ANNs; the
metallicity classifications clearly are of high quality.  Comparison of this
figure with Figure~\ref{fig-teffann} might suggest that the [Fe/H] results are
even superior to the \teff\ ones, despite the large amount of \teff\
information contained in stellar spectra.  This is a matter of appearance only,
as the \teff\ classifications cover a limited range of only 2000~K, {\it i.e.}
a variation of $\leq$~50\% in \teff, while the [Fe/H] classifications cover a
range of $\geq$~3~dex, a factor of more than 1000 in metallicity.

Table~4 summarizes the statistics for the ANNs.  The table is
grouped into three divisions (for the three atmospheric parameters that have
been modeled) of four rows each (for the four subsamples considered, as labeled
in the first column).   For the purposes of making our comparisons, we have
used the robust biweight estimators of central location, $C_{BI}$ (comparable 
to the mean),
and scale, $S_{BI}$ (comparable to the standard deviation), 
as described by Beers, Flynn, \& Gebhardt (1990).  These estimators 
remain resistant to the presence of
outliers, without the need for subjective pruning.  Column (2) of
Table~4 lists the central location of the internal error (the
residual offset in the training sample), in the sense $Q_{ANN} - Q_{CAT}$,
where $Q$ represents the quantity \teff, \logg, and [Fe/H], respectively.
Column (3) is the corresponding central location of the external error (the
residual offset in the testing sample).  Columns (4) and (5) list estimators of
the internal and external scales, respectively.  The final two columns list the
number of stars in the training and testing subsamples, respectively.  Note
that, as expected, the central locations and scales of the internal errors are
generally substantially smaller than those of the external errors.
Nevertheless, the central locations of the external errors are quite
acceptable, and close to zero.

\begin{figure}
\epsscale{0.6}
\plotone{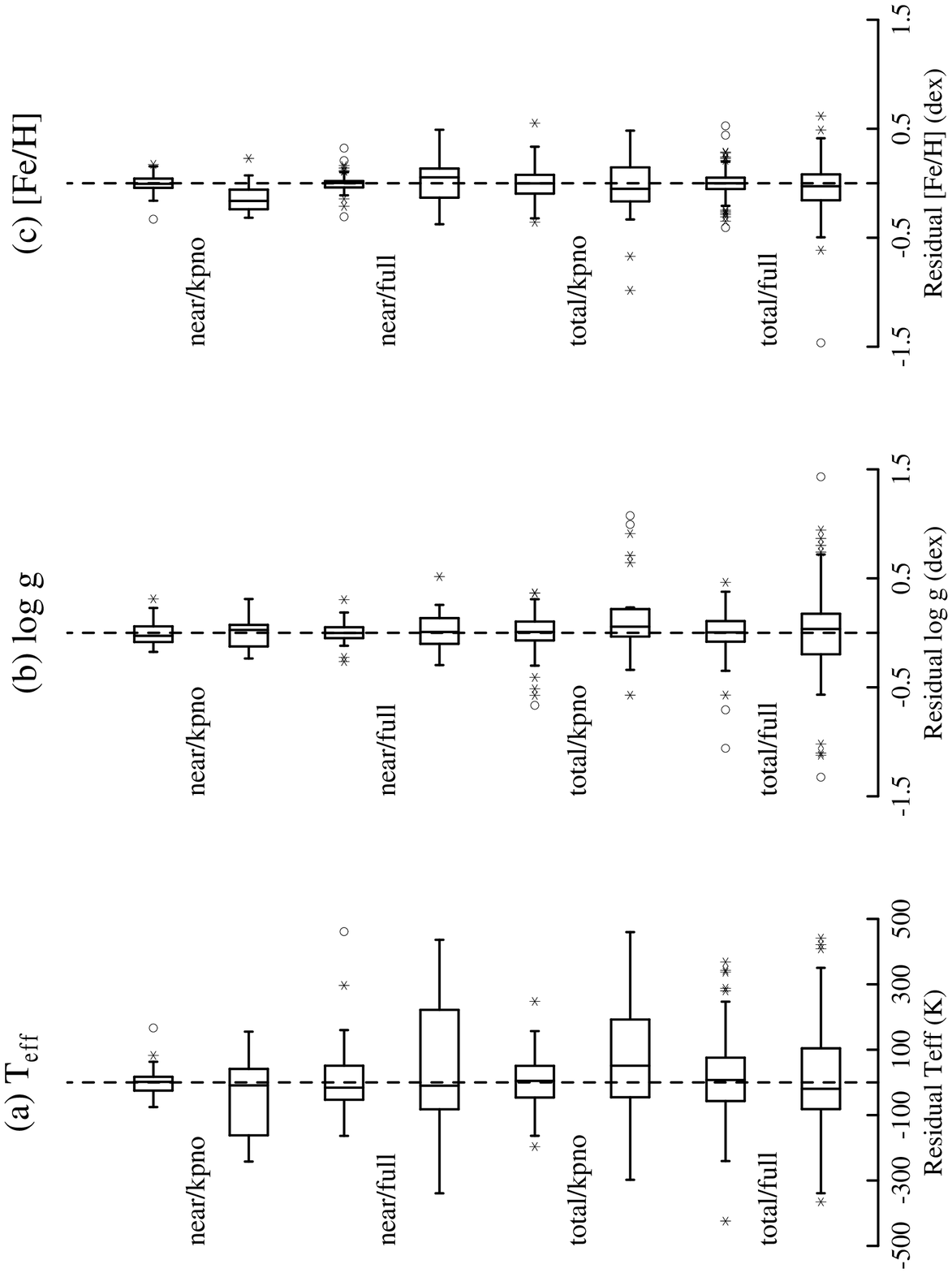}
\figcaption{
Boxplots illustrating comparisons between residuals obtained by the ANNs for
the four data subsamples, in the sense $Q_{ANN} - Q_{CAT}$, where $Q$
represents the quantity \teff\ (panel a), \logg\ (panel b),  and [Fe/H] (panel
c), respectively.   The boxplots immediately above the subsample labels are
those obtained from the training sets, while those immediately below the
subsample labels are those obtained from the testing sets.  The vertical line
in each boxplot is the location of the median residual.  The box extends to
cover the central 50\% of the data.  The ``whiskers'' on each box extend to
cover the last portion of the data not considered likely outliers.  The
asterisks and open circles indicate modest and large outliers, respectively.
\label{fig-tgfres}}
\end{figure}

Figure~\ref{fig-tgfres} is a graphical summary of the distribution of residuals
for the three ANNs, grouped according to the training and testing data.  In
this figure, the training data is shown above the label for each subsample, and
the testing data is shown below the label for each subsample.
The vertical line in each boxplot is the location of the median residual.  The
box extends to cover the central 50\% of the data (the inter-quartile range,
$IQR$).  The ``whiskers'' on each box extend to cover the last portion of the
data not considered likely outliers (this range extends to cover the distance
from the lower and upper ends of the $IQR$ + a factor of $1.5\; \times \; IQR$.  The
asterisks and open circles indicate modest and large outliers (lower and upper
ends of the $IQR$ + a factor of $3.0\; \times \; IQR$), respectively.  See Emerson
\& Strenio (1983) for a general discussion of boxplots.

The scales of the external errors in Table~4 provide our best
estimates of the performance of the ANNs.  The scale estimates obtained for the
total/full subsample for each of the three ANNs, $S_{BI} (\teff) = 185$ K,
$S_{BI} (\logg) = 0.36$ dex, and $S_{BI} ({\rm [Fe/H]}) = 0.21$ dex, are all
acceptably low.  Note that, in general, the scale estimates obtained for the
nearby/kpno subsample are often somewhat smaller than those for the total/full
subsample. We expect this result due to the more homogeneous nature 
and larger spectral coverage of the kpno
data, relative to the full data.  Furthermore, the spectra of the nearby stars
typically have higher S/N than those included in the total/full subsample
and,  in
the case of the \logg\ results, it should be kept in mind that the
trigonometric gravities  for the nearby stars are more
accurate than those obtained for the more distant stars included in the
total/full subsample.

Our external scale errors in the estimates of the atmospheric parameters {\it
include} the errors in the determination of the parameters for the program
stars, i.e., the catalog values.  Given that the catalog values were drawn from
a variety of sources, and no doubt incorporate a number of systematic offsets
from star to star, we conservatively estimate that the errors of determination
for the catalog values are of the order $\sigma (\teff) \sim 100-125$ K, $\sigma
(\logg) \sim 0.20-0.25$ dex, and $\sigma ({\rm [Fe/H]}) \sim 0.10-0.15$ dex,
respectively.  Subtracting these contributions to the external scale estimates
obtained for the total/full subsample suggests that our likely errors in the
physical quantities lie in the range $\sigma (\teff) \sim 135-150$ K, $\sigma
(\logg) \sim 0.25-0.30$ dex, and $\sigma ({\rm [Fe/H]}) \sim 0.15-0.20$ dex,
respectively.  The internal scale errors obtained from inspection of the
training set for the total/full subsample are quite low, $S_{BI} (\teff) = 110$
K, $S_{BI} (\logg) = 0.15$ dex, and $S_{BI} ({\rm [Fe/H]}) = 0.09$ dex,
suggesting that the intrinsic accuracy of our technique is limited by the
accuracy of the training catalog values themselves, and not by any clear
deficiency of the ANN approach.

Katz \etal\ (1998) have pursued a study of a least-squares matching technique,
based on the comparison of high-resolution stellar spectra to a grid of stars
with known atmospheric parameters, and obtained {\it internal} errors of
estimation of $\sigma (\teff) \sim 85-100$ K, $\sigma (\logg) \sim 0.28$ dex, and
$\sigma ({\rm [Fe/H]}) \sim 0.17$ dex, respectively, for stellar spectra with S/N
ratios in the range 10 to 100.  These errors are completely in line with
our own internal errors, suggesting that, when applying the ANN technique, one
is not forced to employ high-resolution spectroscopy, at least for accurate
determination of these stellar parameters.

\subsection{Limitations of the ANN Approach, and Deviations from the General
Trends}

The scatter of points seen in Figure~\ref{fig-tgfres} is dominated by a small
number of stars with large mismatches between input catalog parameters and
output network predictions.  Blame for these clashes must be assessed on a
case-by-case basis, and is provided in Appendix A. Included in that list are
stars with $\mid$ANN-CAT$\mid$ parameter deviations in excess of 350~K in
\teff, 0.6 dex in \logg, and 0.3 dex in [Fe/H].  Below we discuss some factors
that may be responsible when good points go bad.

\begin{figure}
\epsscale{0.5}
\plotone{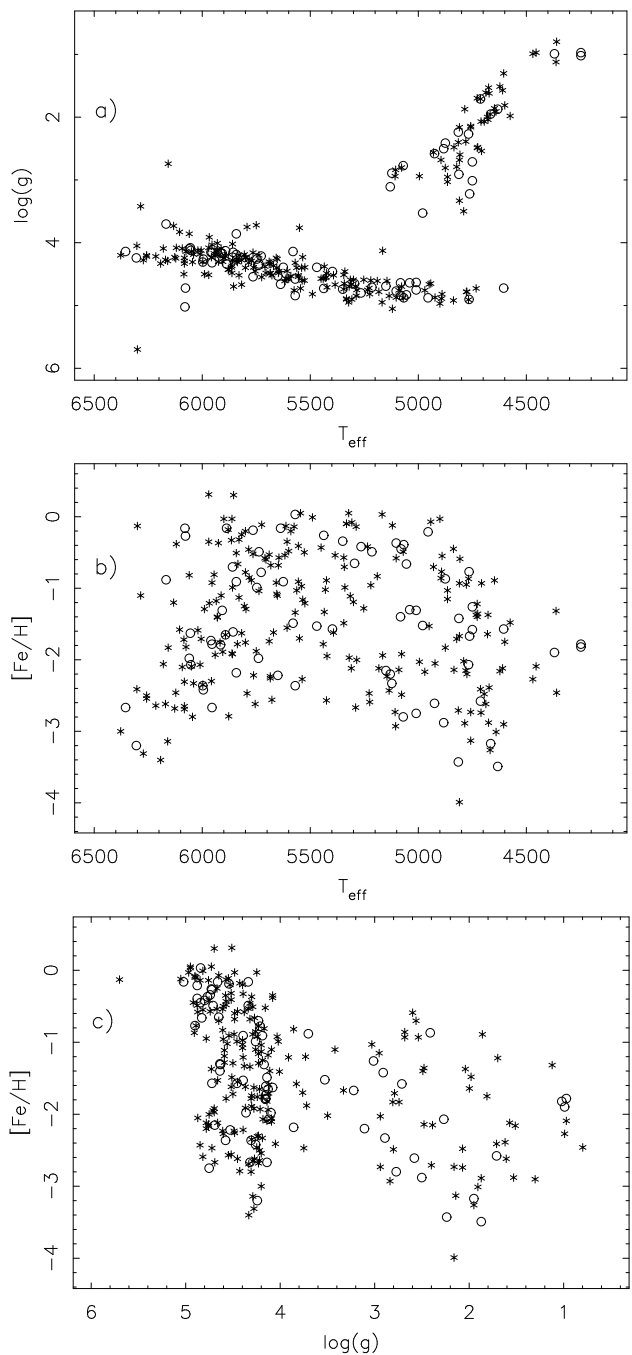}
\figcaption{
Distribution of the catalog atmospheric parameters for the training set (asterisk
symbols) and testing set (open circles).  Panel (a) displays the \teff\ vs.
\logg\ distribution, panel (b) displays the \teff\ vs. [Fe/H] distribution, and
panel (c) displays the \logg\ vs. [Fe/H] distribution.  Note that
the testing set data track similar regions of the physical parameter spaces as
do the training set data.  
\label{fig-paramspace}}
\end{figure}

The reader is cautioned again that the ANNs of this study, in common
with all automated pattern recognition algorithms, are far better {\it
interpolators} than {\it extrapolators}.  Our ANNs have trouble in those areas
of the (\teff, \logg, [Fe/H]) parameter space where the training spectra are
few or  absent.  For example, from inspection of the middle panel of
Figure~\ref{fig-paramspace} one expects difficulties for stars with
[Fe/H]~$\lesssim$ --3, especially over the temperature regime
5000~$<$~\teff~$<$~6000~K, where we have no training or test spectra.  Five of
the deviant stars discussed in Appendix A have extremely low metallicities, and
their spectra at moderate resolution have very few strong atomic features.  The
ANNs undoubtedly are losing some parameter sensitivity in this metallicity
regime, a limitation that should be easily overcome by training on larger
samples of lower metallicity stars.

Subgiants, those stars in the parameter space defined roughly by
5100~$<$~\teff~$<$~5600~K and 3.2~$<$~\logg~$<$~3.8
(Figure~\ref{fig-paramspace}, top panel), are apparently not well
represented among our program stars.  However, this gap may be, at
least in part, tied to our adopted methods of setting catalog
parameters for the program stars.  One virtue of the approach outlined
in \S4 lies in its uniformity -- every star is treated as identically
as possible.  But this demands that some more-or-less arbitrary choices
be made.  For example, the adopted temperature scale is that of Alonso
\etal\ (1996, 1999), which is based upon application of the IRFM.  The
IRFM has very few input assumptions, but does rely on model atmospheres
to predict monochromatic IR fluxes.   On the other hand, the adopted 
gravities are derived from the absolute magnitudes of Beers \etal\ (1999),
based on a visual spectral classification.  Consider as one example the
well-studied subgiant  HD~140283.  Our application of the Alonso
temperature calibration yields \teff\ $_{\rm CAT}$~= 5792 K, but a
glimpse at several recent high-resolution analyses shows a wide range of
values: 5640 K (Magain 1989 $\equiv$ M89), 5750 K (Ryan, Norris, \&
Beers 1996 $\equiv$ RNB96), 5755-5779 (Gratton, Carretta, \& Castelli
1996 $\equiv$ GCC96), and 5843 K (Fuhrmann et al. 1997 $\equiv$
FPFRG97).  This particular case exemplifies a general tendency: the
temperature scales from the IRFM advocated by Alonso \etal\ are neither
among the highest nor the lowest scales in the literature. We assigned
\logg\ $_{\rm CAT} = 3.75$ to this star, close to the  highest values
among the high-dispersion studies: 3.10 (M89), 3.40 (RNB6), 3.60-3.80
(GCC96), and 3.20 (FPFRG97). However, Allende Prieto et al. (1999) derived
$\log g \sim$ 3.80 from the measured {\it Hipparcos} parallax.  Finally,
[Fe/H]$_{\rm CAT} = -2.47$, which falls in the middle of the range spanned by
the high-dispersion analyses: $-2.70$ (M89), $-2.54$ (RNR96), $-2.38$ to
$-2.42$ (GCC96), and $-2.34$ (FPFRG97).

We want to emphasize that considerable care is required in the examination of
the temperature, gravity, and metallicity scales adopted in this and in other
studies.  The same warning applies to individual cases of deviations between
input catalog and output ANN parameters, as not all of our program stars have
been treated with equal vigor in past studies.  Indeed, a few of our program
stars have not yet had the benefit of high-resolution spectroscopic analysis
over wide wavelength ranges.  The ANNs constructed here may fail for particular
stars, but often they also bring to light stars that deserve further study.

\subsection{Experiments on Spectra with Artificially Increased Noise}

If ANNs are to be successfully employed in the analysis of extremely large
spectroscopic data sets (which we anticipate will become available in the near
future), they will need to work reliably on spectra with both high S/N, like
the spectra employed here, as well as with spectra of much lower S/N.  Our data
are not ideally suited to explore the effects of variable S/N values on ANNs,
but we have attempted a few experiments that may point the way toward more
comprehensive efforts in the future.

\begin{figure}
\epsscale{0.8}
\plotone{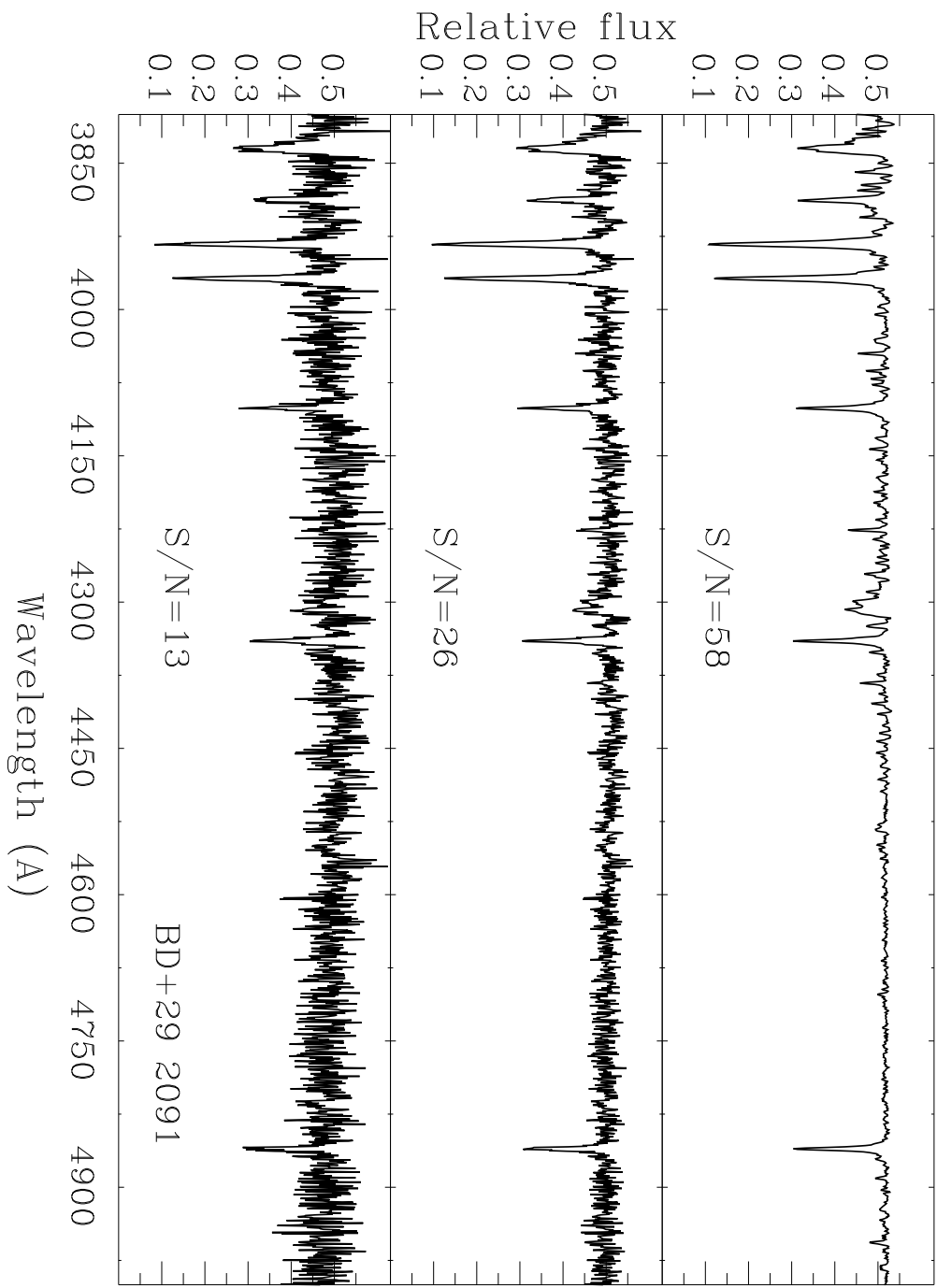}
\figcaption{
The spectrum of BD $+$01 2901, one of the program stars chosen to evaluate the
effect of S/N ratio on the ANN approach.  The top panel shows the original
spectrum, as prepared for submission to the ANNs.  The middle panel shows the
spectrum degraded to a S/N = 26.  The lower panel shows the spectrum degraded
to a S/N = 13.  
\label{fig-noisyspectra}}
\end{figure}

From our original sample of program stars we randomly selected a subset of 52
stars having S/N~$\sim$~50 near 4000~\AA, and artificially degraded their S/N
at this wavelength to $\sim$~26 and then to $\sim$~13.  The strong wavelength
dependence of the S/N in the original spectra was preserved using the square
root of the raw (unflattened) spectra to scale the extra poissonian noise
introduced. In Figure~\ref{fig-noisyspectra} we display one example of the S/N
degradation procedure.  The pernicious effect of low S/N is clearly seen in
this figure, as some prominent features (\eg, the CH G-band near 4300~\AA, or
\ion{Fe}{1} at 4045~\AA) in the original spectrum shown in the top panel become
nearly undetectable in the low S/N spectrum shown in the bottom panel.

As a first experiment, we submitted these sets of spectra with different noise
levels as new data into the final ANNs that we had trained as described above.
In this manner, we sought to ascertain whether or not the ANNs built to
recognize differences in high S/N spectra could produce reasonable estimates of
the atmospheric parameters for stars with lower S/N spectra.

\begin{figure}
\epsscale{0.5}
\plotone{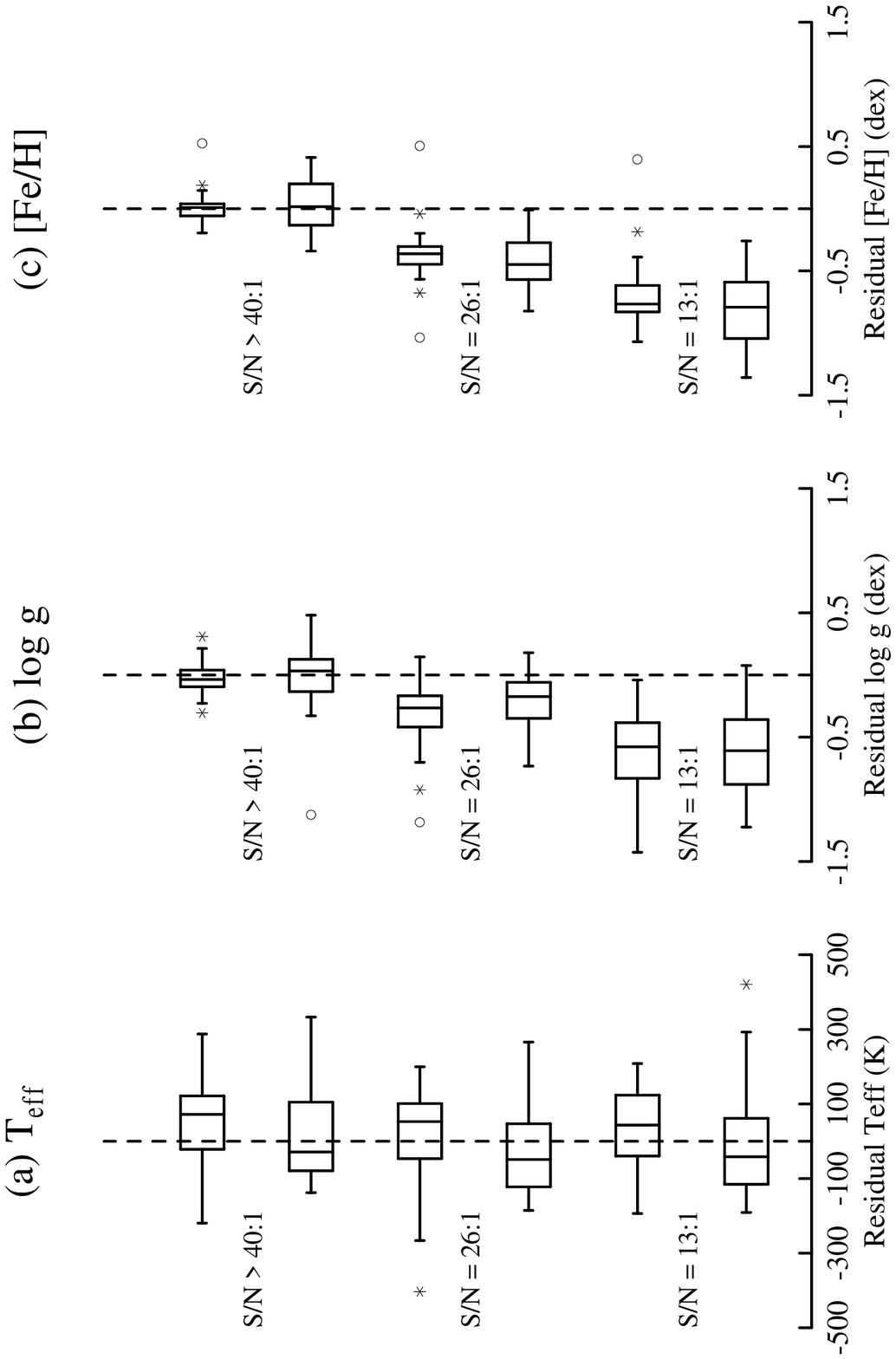}
\figcaption{
Boxplots illustrating comparisons between residuals obtained by the ANNs for
the S/N exploration subsamples, in the sense $Q_{S/N} - Q_{S/N > 40}$, where
$Q$ represents the quantity \teff\ (panel a), \logg\ (panel b), and [Fe/H] (panel
c), respectively.   The boxplots immediately above the subsample labels are
those obtained from the training sets, while those immediately below the
subsample labels are those obtained from the testing sets. Note the obvious
systematic offsets in estimated \logg\ and [Fe/H] for the lower S/N subsamples.
\label{fig-stgfres0}}
\end{figure}

In Table~5 we summarize the statistics of the ANN
classifications of these data sets, organized in a similar manner to
Table~4 above.  Figure~\ref{fig-stgfres0} is a graphical summary
of the distribution of residuals for this experiment.  Inspection of the table
and figure reveal several features of note.  As expected, there is a clear
general trend toward increasing the zero point error ($C_{BI}$) and in the
scatter ($S_{BI}$) for both the internal and external subsamples as one
progresses to lower S/N ratios.  The \teff\ classification is the least
affected, with the changes in location and scale of the residuals staying
almost constant as one progresses from high to low S/N ratios.  The \logg\
classification suffers rapid degradation with declining S/N, and exhibits a
systematic shift in the zero point from near $0.0$ dex to on the order of
$-0.6$ to $-0.7$ dex, and roughly a tripling of the scatter.  Similarly, the
[Fe/H] classification indicates a zero-point shift and increase in scatter with
declining S/N.

The systematic errors in [Fe/H] are puzzling, and opposite those seen in the
auto-correlation function approach described by Beers et al. (1999), where
decreasing S/N leads to a positive systematic error in the metallicity scale.
Furthermore, there also exists a trend in metallicity at a {\it fixed} S/N
ratio, in the sense that metal-rich stars have their abundances more
underestimated than the metal-deficient stars.  At present, we cannot
explain these systematic trends in \logg\ or [Fe/H] classification with
decreasing S/N, and we leave this to future investigation.

\begin{figure}
\epsscale{0.5}
\plotone{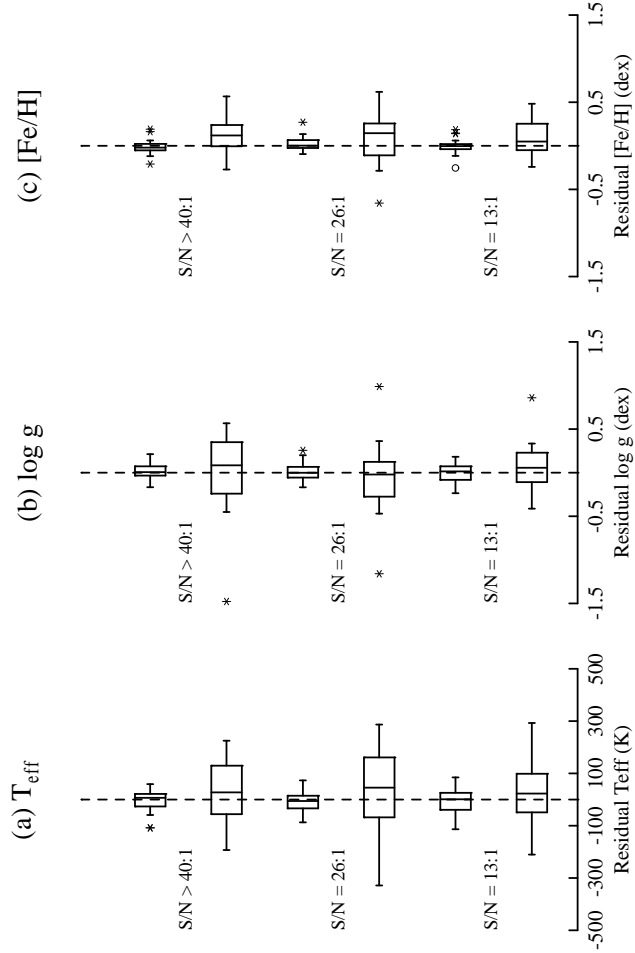}
\figcaption{
Boxplots illustrating comparisons between residuals obtained by the ANNs for
the S/N exploration subsamples, for \teff (panel a), \logg (panel b), and
[Fe/H] (panel c), respectively.  The boxplots immediately above the subsample
labels are those obtained from the training sets, while those immediately below
the subsample labels are those obtained from the testing sets. Note that in
this case, where the ANNs are trained on spectra of similar S/N ratios as that
of the spectra which are submitted to them for evaluation, the systematic
offsets seen in Figure~\ref{fig-stgfres0} disappear.
\label{fig-stgfres}}
\end{figure}

As a final experiment, we constructed {\it new} ANNs, trained on input
spectra having similar S/N ratios to the spectra we test them with, \ie\, S/N
$> 40$, $26$, and $13$ respectively.  Table~5 and
Figure~\ref{fig-stgfres} show the results.  It is immediately clear that the
zero-point shifts previously encountered have now disappeared.  Of even greater
interest, the scatter obtained in the estimates of the internal and external
errors are essentially identical for spectra of declining S/N ratios, and are
roughly equivalent to those obtained previously when using ANNs trained and
tested with only high S/N spectra, as can be noted by comparison of
Figure~\ref{fig-stgfres0} with Figure~\ref{fig-tgfres}.  We conclude that it is
better to classify spectra of a given S/N with ANNs trained on spectra of
similar S/N than with ANNs trained exclusively on higher S/N spectra.  This
result suggests two future approaches.  One should either train ANNs on spectra
with a variety of S/N ratios (and much larger training sets), or train ANNs for
specific S/N ratios, and carry out an interpolation between the derived results
for the observed S/N of the spectrum.

\section{Conclusions}

We have explored the use of artificial neural networks for three-dimensional
classification of medium-resolution stellar spectra.  We have constructed,
trained, and tested ANNs specific to the individual estimation of the \teff,
\logg, and [Fe/H], and find that these parameter-specific networks are superior
to the simultaneous estimation of multiple parameters from an omnibus ANN.  The
external accuracy of the physical parameter estimates, $\sigma (\teff) = 135-150$
K over the range $4250 \le \teff \le 6500$ K, $\sigma (\logg) = 0.25-0.30$ dex
over the range $1.0 \le \logg \le 5.0$ dex, and $\sigma ({\rm [Fe/H]}) =
0.15-0.20$ dex over the range $\-4.0 \le {\rm [Fe/H]} \le +0.3$, strongly
encourage further refinement of this approach for future work.  Furthermore, we
find that the derived accuracies of parameter estimates are not severely
affected by the presence of modest spectral noise, at least when networks are
trained with spectra with similar S/N ratios to those that will be analyzed.
Further experimentation is necessary to identify the limiting S/N ratio for
which useful parameter estimation is still possible; already, we find
reasonably accurate estimates can be obtained with spectra of S/N as low as
13.

In the near future, we anticipate the construction of trained ANNs, covering a
variety of S/N ratios and spectral resolutions, with which stellar spectra
exhibiting a wide range of atmospheric parameters can be usefully analyzed. 
The recent study by Gray, Graham \& Hoyt (2001), which makes use of stellar
spectra very similar to those employed here, has revealed that micro-turbulence 
 has to be taken into account as an independent parameter in order
  to recover properly  the  surface gravity. A
second addition that will improve the results is to decouple the abundances
of the alpha elements from the rest of the metals, modeling it as a new 
variable. So far, the simple tests carried out in this paper  encourage
the use and refinement of ANNs for on-going and 
soon-to-be-undertaken large-scale surveys of 
stellar spectra, both from ground-based and space-based observatories.

\acknowledgments

We thank B.D. Ripley for the use of his neural network software and A.
Alonso for providing his photometric calibrations prior to publication.

We gratefully acknowledge major financial support for this work from the State
of Texas Advanced Research Program.  This research was funded in part by NSF
grants AST-0086321 to CAP, AST-9618364 to CS, and AST-9529454 to TCB.
TvH acknowledges support from the Gemini Observatory, which is operated by
AURA, Inc., under a cooperative agreement with the NSF on behalf of the
Gemini partnership:  the NSF (United States), PPARC (United Kingdom), NRC
(Canada), CONICYT (Chile), ARC (Australia), CNPq (Brazil) and CONICET
(Argentina).  SR acknowledges partial support for this research from
FAPESP (98/02706-6) and from CNPq.

This work made use of the SIMBAD database, maintained at the CDS, and
NASA's ADS.

\appendix

\section{Comments on Deviant Stars}

\begin{figure}
\epsscale{0.7}
\plotone{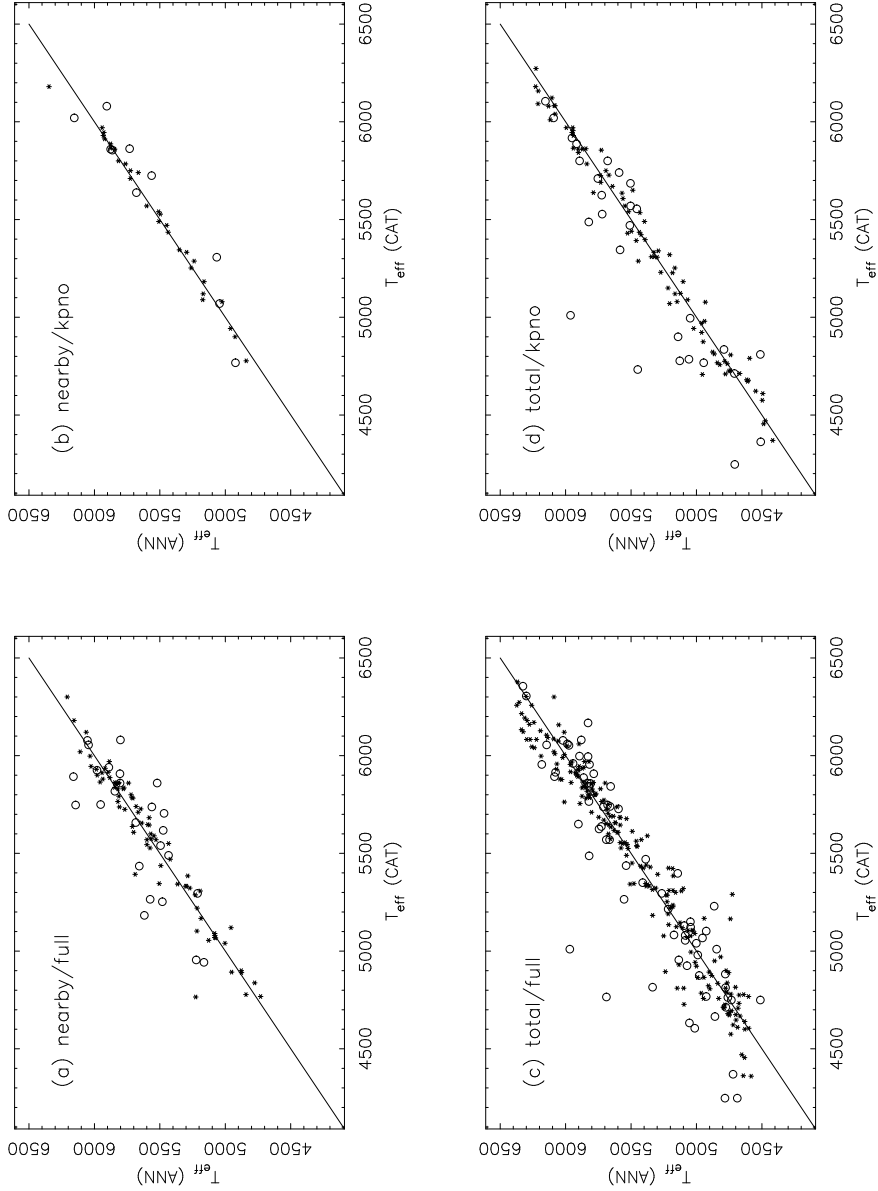}
\figcaption{
Results obtained by the \teff\ ANNs for the four data subsamples.  See the text
for the definition of the subsamples.  The horizontal axes display the catalog
(CAT) \teff\ values, while the vertical axes display the network classification
(ANN) \teff\ values, both in Kelvins.  The line drawn in each panel is the
one-to-one correspondence line.  The points plotted as asterisk symbols are for
the training set, and the points plotted as the open circle symbols are for the
testing set.  
\label{fig-teffann}}
\end{figure}

\begin{figure}
\epsscale{0.8}
\plotone{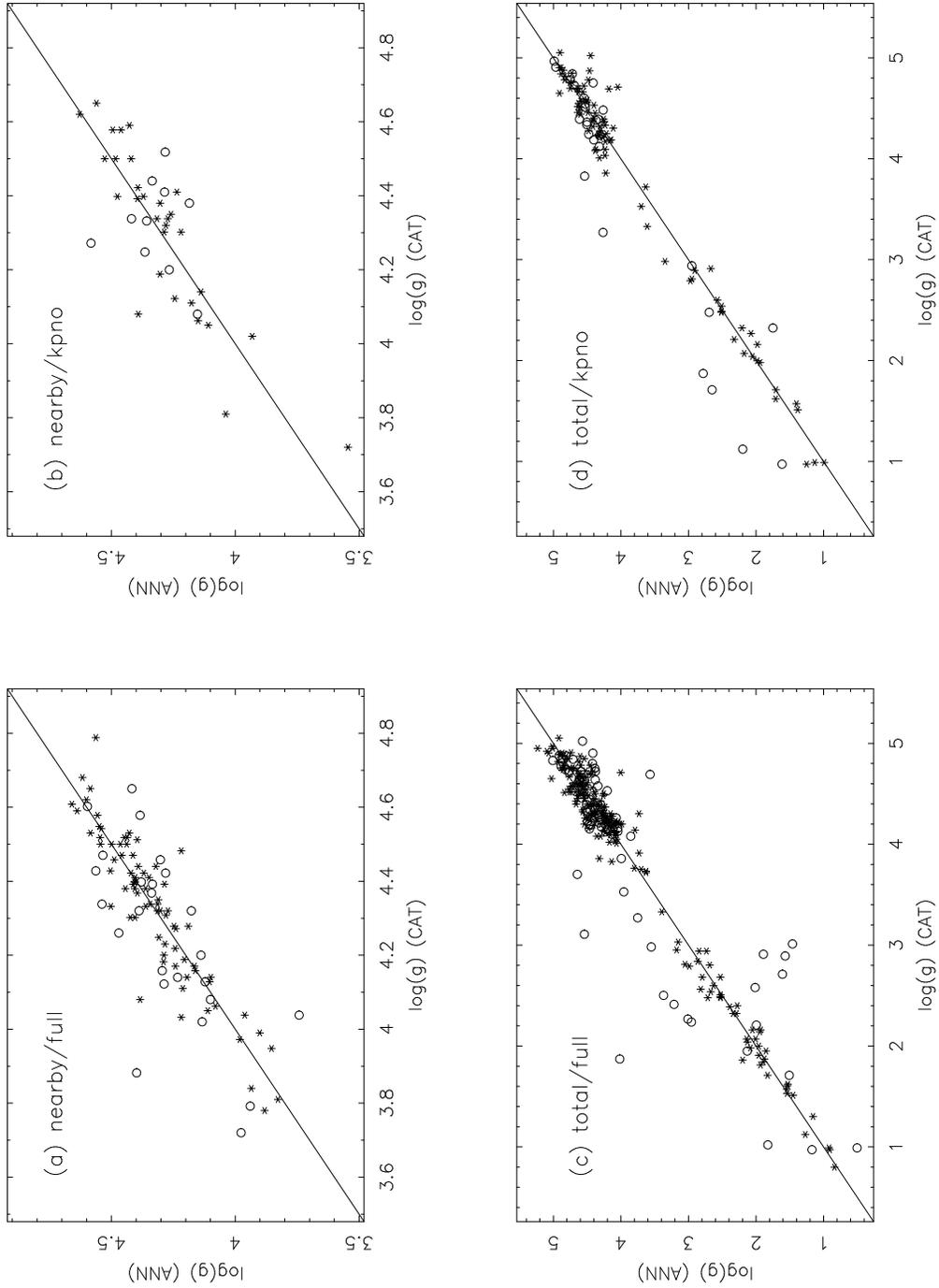}
\figcaption{
Results obtained by the \logg\ ANNs for the four data subsamples.  Note that
the axes of panels (a) and (b) display a much smaller range of \logg\ values
than the axes of panels (c) and (d). Lines and symbols are as in
Figure~\ref{fig-teffann}.
\label{fig-loggann}}
\end{figure}

\begin{figure}
\epsscale{0.8}
\plotone{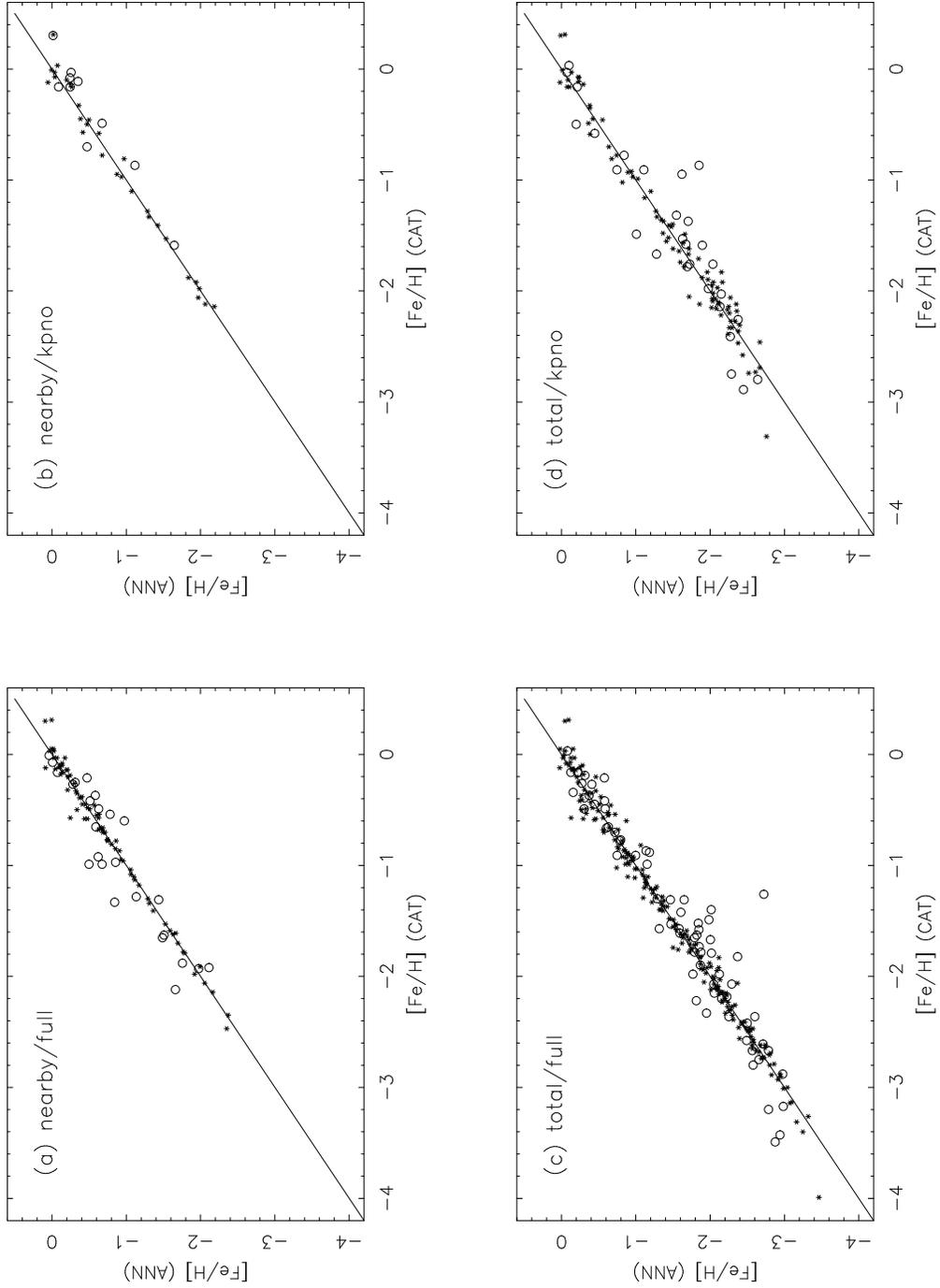}
\figcaption{
Results obtained by the [Fe/H] ANNs for the four data subsamples.  Lines and
symbols are as in Figure~\ref{fig-teffann}.  
\label{fig-fehann}}
\end{figure}

In spite of the excellent general predictive capability of our ANNs, some stars
obviously have poor matches between one or more of their CAT and ANN atmospheric
parameters.  In this appendix we draw attention to those cases with parameter
clashes $\mid$CAT--ANN$\mid$ that are larger than 350~K in \teff, 0.6 dex in
\logg, and  0.3 dex in [Fe/H].  The discrepant cases are noted in
Table~2  and Table~3 with a colon next to the
relevant parameters. Since one expects (and we found) a larger overall
agreement for the training set than for the testing set, these subsets will be
considered separately.  We refer the reader back to
Figures~\ref{fig-teffann}--\ref{fig-fehann} to visually locate these deviant
stars in relation to the vast majority of conforming stars.

For each star only the deviating parameters will be quoted here; see
Tables~2  and 3 for the remaining parameters.
Comments given here will primarily address comparisons with previous papers
that report extensive abundance analyses from high-resolution spectroscopy.
Several useful papers discuss the derivation of atmospheric parameters for large
stellar samples from medium-resolution ($R \equiv \lambda$/$\Delta\lambda
\sim$~5000) spectra (\eg, Beers \etal\ 1999; Ryan \& Norris 1991), or from very
low S/N, small wavelength-coverage spectra (\eg, Carney \etal\ 1994, and
references therein).  Such studies have formed the basis for compilation of our
catalog parameters, so will not be re-examined in detail here.  In citing
literature sources to support catalog or ANN parameters, it should be
understood that although the results of various studies are hopefully
internally self-consistent, in our application they were normalized to a
variety of \teff, \logg, and [Fe/H] systems.  Consequently, the reader is urged
to view the following comments with indulgence.

First we consider the parameter mismatches in the ANN training set.  Two stars
of this set have not been studied extensively with high-spectral resolution
data:  G 66-49 ([Fe/H]$_{\rm CAT}$~= --0.57, [Fe/H]$_{\rm ANN}$~= --0.13), and
G 236-11 ([Fe/H]$_{\rm CAT}$~= +0.31, [Fe/H]$_{\rm ANN}$~= --0.10).  Note that
the Beers \etal\ (1999) medium-resolution study obtained estimates of
metallicity of [Fe/H]$_{\rm AK2} = -0.32$ for G 66-49 and [Fe/H]$_{\rm AK2} =
-0.20$ for G 236-11, closer to the predictions of the ANN.  These two stars
will not be discussed further here.  For the handful of other discrepant stars,
we list below a few brief comparisons to the literature.

\begin{itemize}

\item BD $+$37 1458 (\logg\ $_{\rm CAT}$~= 4.71, \logg\ $_{\rm ANN}$~= 4.00):
This star's {\it Hipparcos} parallax (ESA 1997) is consistent with subgiant
evolutionary status, and Gratton \etal\ (2000) derive \logg\ = 3.3, thus
the ANN gravity value is clearly to be preferred over that of the catalog.

\item CS 22949-037 ([Fe/H]$_{\rm CAT}$~= --3.99, [Fe/H]$_{\rm ANN}$~= --3.46):
This extremely metal-poor star is warm enough to have little heavy-element line
absorption at moderate spectral resolution.  The ANN can recognize the star's
low metallicity, but cannot be expected to derive a very accurate abundance
estimate due to the few training stars at such low metallicities and
intermediate temperatures.  Nevertheless, comparison with the recent
high-resolution analysis of Norris, Ryan, \& Beers (2001), who obtain [Fe/H] =
--3.79, indicates that the correct abundance lies roughly halfway between the
CAT and ANN values.  Note that the Beers \etal\ (1999) abundance, [Fe/H]$_{\rm
AK2} = -3.46$, exactly matches the ANN determination.

\item G 58-30 = HD~94835 ([Fe/H]$_{\rm CAT}$~= +0.30, [Fe/H]$_{\rm ANN}$~=
--0.05): Feltzing \& Gustafsson (1998) derive [Fe/H]~= +0.13, splitting the
difference between the catalog and ANN values.  The Beers \etal\ (1999)
abundance estimate for this star is [Fe/H]$_{\rm AK2} = -0.32$, closer to the
ANN value.

\item HD 84937 ([Fe/H]$_{\rm CAT}$~= --2.06, [Fe/H]$_{\rm ANN}$~= --2.37):
Carretta, Gratton, \& Sneden (2000) derive [Fe/H]~= --2.04 from a reanalysis of
literature data; the approximate mean value of other recent literature sources
(Cayrel de Strobel \etal\ 1997) suggests [Fe/H]~= --2.2, so the catalog
metallicity is probably to be preferred.  The Beers \etal\ (1999) estimate for
HD 84937 is [Fe/H]$_{\rm AK2} = -2.14$, closer to the catalog abundance
estimate.

\item HD~105546 (\teff\ $_{\rm CAT}$ = 4727~K, \teff\ $_{\rm ANN}$~= 5095~K):
The ANN temperature is in better agreement with high-resolution spectroscopic
studies: \teff~= 5300~K (Pilachowski, Sneden, \& Kraft 1996), and \teff~=
5147~K (Gratton \etal\ 2000).

\item HD~136202 (\logg\ $_{\rm CAT}$~= 5.70, \logg\ $_{\rm ANN}$~= 4.64): The
latest update of the Cayrel de Strobel \etal\ (1997) catalog lists literature
studies deriving $<$\logg$>$~= 4.0.  The SIMBAD database lists a spectral type
of F8 III-IV.  Therefore the inferred extremely high catalog \logg\ is
incorrect.

\item HD~218857 (\teff\ $_{\rm CAT}$ = 5165~K, \teff\ $_{\rm ANN}$~= 4740~K):
Previous high-resolution analyses (\eg, \teff~= 5125~K, Pilachowski 
\etal\ 1996) and $B-V$~= 0.65 from the SIMBAD database support the higher
catalog temperature.

\item LP 685-44 (\teff\ $_{\rm CAT}$ = 5290~K, \teff\ $_{\rm ANN}$~= 4726~K):
No extensive analysis of this star using high spectral resolution
data has been published.  The SIMBAD database has $B-V$~= 0.63, consistent
with the higher catalog \teff.

\end{itemize}

Next we consider parameter estimation problems in the testing set of stars.
The discrepant stars of this set that apparently lack extensive high-resolution
spectroscopic analyses are:

G 17-22 = HD 149162 (\teff\ $_{\rm CAT}$ = 4765~K, \teff\ $_{\rm ANN}$~=
5687~K); G 99-52 ([Fe/H]$_{\rm CAT}$~= --1.40, [Fe/H]$_{\rm ANN}$~= --2.01); G
106-53 ([Fe/H]$_{\rm CAT}$~= --0.21, [Fe/H]$_{\rm ANN}$~= --0.58); G 146-76
(\logg\ $_{\rm CAT}$~= 4.69, \logg\ $_{\rm ANN}$~= 3.57); and G 161-84 (\teff\
$_{\rm CAT}$ = 4605~K, \teff\ $_{\rm ANN}$~= 5013~K).  The Beers \etal\ (1999)
abundance determination for G 99-52 is [Fe/H]$_{\rm AK2} = -0.87$, while that
for G 106-53 is  [Fe/H]$_{\rm AK2} = -0.31$.  These stars will not be discussed
further here.  Below we list comments based on comparisons with the literature
for this somewhat larger list of discrepant stars.

\begin{itemize}

\item BD $+$01 2916 (\teff\ $_{\rm CAT}$ = 4247~K, \teff\ $_{\rm ANN}$~= 4782~K;
\logg\ $_{\rm CAT}$~= 1.02, \logg\ $_{\rm ANN}$~= 1.83; 
[Fe/H]$_{\rm CAT}$~= --1.82, [Fe/H]$_{\rm ANN}$~= --2.37): High-resolution
studies (Cayrel de Strobel \etal\ 1997, Shetrone 1996) favor the catalog values
of all parameters. The Beers \etal\ (1999) abundance determination for this
star, [Fe/H]$_{\rm AK2} = -1.60$, seems to support the catalog value as well.

\item BD $-$04 680 (\teff\ $_{\rm CAT}$ = 5650~K, \teff\ $_{\rm ANN}$~= 5902~K;
[Fe/H]$_{\rm CAT}$~= --2.22, [Fe/H]$_{\rm ANN}$~= --1.81):
Bonifacio \& Molaro (1997) recommend \teff~= 5866~K, \logg~=3.73, and [Fe/H]~=
--2.07, in rough agreement with the means of the CAT and ANN temperatures and
metallicities, but their gravity value is much lower than either of our
estimates, so further investigation of this star is warranted. The Beers \etal\
(1999) abundance determination for this star, [Fe/H]$_{\rm AK2} = -2.17$,
agrees well with the catalog estimate.

\item BD $-$14 5890 (\logg\ $_{\rm CAT}$~= 2.27, \logg\ $_{\rm ANN}$~= 3.01):
The analysis of Bonifacio, Centurion, \& Molaro (1999), drawing on results of
an earlier study by Cavallo, Pilachowski, \& Rebolo (1997), yields gravity
estimates of \logg~= 2.34 from the star's {\it Hipparcos parallax}, and
\logg~ = 1.4 from a spectrum analysis; these appear to rule out the higher ANN
gravity.  Note also that Bonifacio \etal\ derived [Fe/H]~= --2.52,
substantially lower than either of our metallicity estimates.  The Beers \etal\
(1999) abundance determination for this star is [Fe/H]$_{\rm AK2} = -2.07$,
midway between the CAT and ANN values, and again, rather different from the
Bonifacio \etal\ estimate.

\item CS~22873-128 (\logg\ $_{\rm CAT}$~= 2.50, \logg\ $_{\rm ANN}$~= 3.37):
McWilliam \etal\ (1995) derive \logg\ ~= 2.1 for this extremely metal-poor
giant, and our ANN probably does not have many good \logg\ indicators in this
cool star's very weak-lined spectrum.

\item CS~22891-200 (\teff\ $_{\rm CAT}$ = 4632~K, \teff\ $_{\rm ANN}$~= 5053~K;
\logg\ $_{\rm CAT}$~= 1.87, \logg\ $_{\rm ANN}$~= 4.02; [Fe/H]$_{\rm CAT}$~=
--3.49, [Fe/H]$_{\rm ANN}$~= --2.88): The McWilliam \etal\ (1995)
high-dispersion analysis provides a temperature estimate, $\teff = 4700$~K,
that matches the catalog value, but differs somewhat in its derived surface
gravity estimate, $\logg = 1.0$.  The ANN \logg\ is clearly incorrect.

\item CS~22968-014 (\teff\ $_{\rm CAT}$ = 4815~K, \teff\ $_{\rm ANN}$~= 5335~K;
\logg\ $_{\rm CAT}$~= 2.24, \logg\ $_{\rm ANN}$~= 2.96;
[Fe/H]$_{\rm CAT}$~= --3.43, [Fe/H]$_{\rm ANN}$~= --2.94): The McWilliam \etal\
(1995) high-dispersion analysis completely supports the catalog values for this
star --  yet another example of our trained ANNs having trouble with a cool,
very weak-lined spectrum.  The Beers \etal\ (1999) abundance determination for
this star, [Fe/H]$_{\rm AK2} = -3.35$, is closer to the catalog value.

\item G 21-22 (\teff\ $_{\rm CAT}$ = 6167~K, \teff\ $_{\rm ANN}$~= 5828~K;
\logg\ $_{\rm CAT}$~= 3.70, \logg\ $_{\rm ANN}$~= 4.64;
[Fe/H]$_{\rm CAT}$~= --0.88, [Fe/H]$_{\rm ANN}$~= --1.18): Bonifacio \& Molaro
(1997) derive \teff~= 5869~K, \logg~= 3.93, and [Fe/H]~= --1.63, thus agreeing
with the ANN result for temperature, with the catalog input for gravity, and
with neither for metallicity!  The Beers \etal\ (1999) abundance determination
for G 99-52 is [Fe/H]$_{\rm AK2} = -0.87$, in agreement with the catalog
estimate.  This obviously is a case for further exploration on all fronts.

\item HD~6755 (\teff\ $_{\rm CAT}$ = 5230~K, \teff\ $_{\rm ANN}$~= 4864~K;
[Fe/H]$_{\rm CAT}$~= --1.49, [Fe/H]$_{\rm ANN}$~= --1.98):  All of the
literature (Cayrel de Strobel \etal\ 1997) studies support the catalog values
for this star, as does the Beers \etal\ (1999) abundance determination,
[Fe/H]$_{\rm AK2} = -1.35$; the ANN result is clearly in error.

\item HD~20038 (\logg\ $_{\rm CAT}$~= 2.41, \logg\ $_{\rm ANN}$~= 3.21):
Gratton \etal\ (2000) derive \logg~= 2.38, so the catalog value is to be preferred.

\item HD~44007 (\logg\ $_{\rm CAT}$~= 2.71, \logg\ $_{\rm ANN}$~= 1.61):
The mean of the entries in Cayrel de Strobel \etal\ (1997) suggest
\logg~$\simeq$~2.1, nearly splitting the difference between the CAT and ANN
values.

\item HD~74462 (\logg\ $_{\rm CAT}$~= 2.91, \logg\ $_{\rm ANN}$~= 1.89): Most
previous studies (Cayrel de Strobel \etal\ 1997) support the ANN value, and
Gratton \etal\ (2000) derive \logg~=1.56, thus the catalog entry appears
incorrect.

\item HD~111721 (\logg\ $_{\rm CAT}$~= 3.01, \logg\ $_{\rm ANN}$~= 1.46;
[Fe/H]$_{\rm CAT}$~= --1.26, [Fe/H]$_{\rm ANN}$~= --2.72): The high-resolution
analysis of Gratton \etal\ (2000) obtains \logg~= 2.5 and [Fe/H]~=--1.27, in
support of the catalog values.  The Beers \etal\ (1999) abundance determination
for this star is [Fe/H]$_{\rm AK2} = -0.88$, also in better agreement with the
catalog value.

\item HD~128279 (\logg\ $_{\rm CAT}$~= 3.11, \logg\ $_{\rm ANN}$~= 4.54):
This star is clearly evolved from the main sequence, as Pilachowski \etal\
(1996) derive \logg~=2.8, and Gratton \etal\ (2000) obtain a value of 3.0.  The
catalog gravity is correct.

\item HD~187111 (\teff\ $_{\rm CAT}$ = 4247~K, \teff\ $_{\rm ANN}$~= 4688~K):
Most literature sources (Cayrel de Strobel \etal\ 1997) agree with the lower
catalog temperature, but Gratton \etal\ (2000) derive \teff~=4429, in the
middle of the CAT / ANN range.

\item HD~195636: Although the CAT and ANN parameters do not disagree enough to
qualify as discrepant here, we note that this unique star has been described by
Preston (1997) as a star near the transition region between the horizontal and
asymptotic giant branches.  It also is rapidly rotating for a highly evolved
star: v$sin i$~= 25~\kmsec.  Its line spectrum will not fit easily into
classification schemes based on more normal metal-poor stellar spectra.

\item HD~196944 (\logg\ $_{\rm CAT}$~= 2.89, \logg\ $_{\rm ANN}$~= 1.57;
[Fe/H]$_{\rm CAT}$~= --2.33, [Fe/H]$_{\rm ANN}$~= --1.95): This is a
carbon-rich metal-poor star, and our ANNs have not been properly
trained to deal with spectra of those objects.  But the detailed analysis of
Zacs, Nissen, \& Schuster (1998) yields \logg~=1.7 and [Fe/H]~= --2.45,
strongly supporting the ANN gravity while in closer agreement with the catalog 
[Fe/H] value.

\item LP 815-43 ([Fe/H]$_{\rm CAT}$~= --3.20, [Fe/H]$_{\rm ANN}$~= --2.79)
Ryan, Norris \& Bessel (1991) obtain [Fe/H]~= --3.20,
and all agree that the star is a warm main-sequence star; its very
weak-lined spectrum is clearly difficult for our ANN to treat properly 
for metallicity.  However, the Beers \etal\ (1999) abundance determination for
this star is [Fe/H]$_{\rm AK2} = -2.92$, closer to the ANN value.

\item Ross-740 = LTT 743 (\teff\ $_{\rm CAT}$ = 5010~K, \teff\ $_{\rm ANN}$~=
5968~K):  Ryan, \etal\ (1991) derive \teff~=5500, \logg~=3.2, [Fe/H]~= --2.75,
and Beers \etal\ (1999) obtain [Fe/H]$_{\rm AK2} = -2.91$, so both of our
metallicity estimates appear to be reasonable, but both the catalog and the ANN
claim main-sequence gravity for a star that seems to be a subgiant from
high-resolution analysis.

\end{itemize}

\clearpage


\begin{deluxetable}{llllr}
\rotate
\label{tab-datasets}
\scriptsize
\tighten
\tablenum{1}
\tablewidth{7.5in}
\tablecaption{The Spectroscopic Data Sets}
\tablecolumns{5}
\tablehead{
\colhead{Telescope}                        &
\colhead{Detector}                         &
\colhead{Coverage}                         &
\colhead{Dispersion}                       &
\colhead{Number}                           \\
\colhead{}                                 &
\colhead{}                                 &
\colhead{(\AA)}                            &
\colhead{(\AA/pix)}                        &
\colhead{}                                 
}
\startdata
ESO 1.5m              & Ford + Loral 2048$\times$2048  & 3750$-$4750  & 0.65 + 0.50&  52 \\
                       &                                &              &            &     \\
KPNO 2.1m             & Tek 2048$\times$2048           & 3750$-$5000  & 0.65       & 115 \\
                       &                                &              &            &     \\
LCO 2.5m     & Reticon + 2D-Frutti            & 3700$-$4500  & 0.65       &  50 \\
                       &                                &              &            &     \\
ORM INT 2.5m  & Tek 1024$\times$1024           & 3750$-$4700  & 0.85       &  3  \\
                       &                                &              &            &     \\
PAL 5m            & Reticon + 2D-Frutti            & 3700$-$4500  & 0.65       &  3  \\
                       &                                &              &            &     \\
SSO 2.3m    & SITe 1752$\times$532           & 3750$-$4600  & 0.50       &  58 \\
\enddata
\tablecomments{
ESO $\equiv$ European Southern Observatory; KPNO $\equiv$ Kitt Peak National Observatory; LCO $\equiv$ Las Campanas Observatory; ORM $\equiv$ Observatorio del Roque de los Muchachos; PAL $\equiv$ Palomar Observatory; SSO $\equiv$ Siding Spring Observatory
}
\end{deluxetable}


\begin{deluxetable}{llcccccc} 
\tablenum{2}
\tablecolumns{8}
\tablewidth{0pt}
\tablecaption{Catalog and ANN Parameters for the Training Sample 
\label{tab-trainlist}}
\tablehead{
\colhead{Star}                    & 
\colhead{Source}                  &
\colhead{T$_{\rm eff~ CAT}$} & 
\colhead{T$_{\rm eff~ ANN}$} & 
\colhead{$\log g_{\rm CAT}$}      &  
\colhead{$\log g_{\rm ANN}$}      & 
\colhead{[Fe/H]$_{\rm CAT}$}      & 
\colhead{[Fe/H]$_{\rm ANN}$}      \\
\colhead{}                        & 
\colhead{}                        &
\colhead{(K)}                     & 
\colhead{(K)}                     & 
\colhead{(dex)}                   & 
\colhead{(dex)}                   & 
\colhead{(dex)}                   & 
\colhead{(dex)}  
}
\startdata 
BD +03 2782     &  K    &  4790  &  4770 &	 2.32 &  2.29 &	 $-$2.02& $-$2.05  \\
BD +04 2621     &  K    &  4712  &  4795 &	 1.71 &  1.83 &	 $-$2.41& $-$2.43  \\
BD +06 648      &  K    &  4455  &  4633 &	 0.97 &  0.91 &	 $-$2.09& $-$2.11  \\
BD +09 2870     &  K    &  4672  &  4749 &	 1.62 &  1.53 &	 $-$2.39& $-$2.31  \\
BD +09 352      &  E    &  6050  &  5987 &	 4.14 &  4.25 &	 $-$2.09& $-$2.09  \\
BD +10 2495     &  K    &  4875  &  4884 &	 2.81 &  3.05 &	 $-$1.83& $-$2.12  \\
BD +17 3248     &  K    &  4995  &  5042 &	 2.94 &  2.73 &	 $-$2.03& $-$2.13  \\
BD +17 4708     &  O    &  6085  &  6092 &	 4.50 &  4.25 &	 $-$1.72& $-$1.79  \\
BD +19 1185     &  K *  &  5435  &  5427 &	 4.38 &  4.31 &	 $-$1.33& $-$1.22  \\
BD +30 2611     &  K    &  4362  &  4642 &	 1.12 &  1.27 &	 $-$1.32& $-$1.38  \\
BD +37 1458     &  K    &  5422  &  5181 &	 4.71: &  4.00: & $-$1.95& $-$2.09  \\
BD $-$01 2582   &  E    &  5145  &  5125 &	 4.61 &  4.71 &	 $-$2.23& $-$2.16  \\
BD $-$13 3442   &  E    &  6160  &  6276 &	 4.29 &  4.28 &	 $-$3.14& $-$3.08  \\
BD $-$18 5550   &  K    &  4785  &  4963 &	 1.87 &  1.88 &	 $-$2.89& $-$2.83  \\
CD $-$31 622    &  L    &  5285  &  5224 &	 4.75 &  4.68 &	 $-$2.00& $-$2.04  \\
CD $-$33 3337   &  L    &  5930  &  6069 &	 4.11 &  4.09 &	 $-$1.40& $-$1.32  \\
CD $-$71 1234   &  E    &  6082  &  6297 &	 4.29 &  4.40 &	 $-$2.65& $-$2.57  \\
CS 22873-055    &  L    &  4675  &  4700 &	 1.53 &  1.54 &	 $-$2.88& $-$2.94  \\
CS 22873-166    &  L    &  4605  &  4600 &	 1.30 &  1.16 &	 $-$2.90& $-$2.95  \\
CS 22878-101    &  E    &  4757  &  4947 &	 2.14 &  1.94 &	 $-$3.13& $-$3.10  \\
CS 22892-052    &  E    &  4640  &  4632 &	 1.91 &  1.96 &	 $-$3.01& $-$2.99  \\
CS 22896-154    &  L    &  5107  &  5032 &	 2.94 &  2.84 &	 $-$2.73& $-$2.72  \\
CS 22947-187    &  E    &  5077  &  5251 &	 2.80 &  2.69 &	 $-$2.49& $-$2.50  \\
CS 22949-037    &  P    &  4810: &  5097:&	 2.16 &  2.05 &	 $-$3.99:& $-$3.46:  \\
CS 22952-015    &  P    &  4667  &  4666 &	 1.95 &  1.85 &	 $-$3.26& $-$3.32  \\
G 5-19          &  K    &  5607  &  5606 &	 4.45 &  4.54 &	 $-$1.55& $-$1.50  \\
G 8-16          &  K *  &  6020  &  6089 &	 4.12 &  4.07 &	 $-$1.59& $-$1.68  \\
G 9-27          &  K    &  5440  &  5383 &	 4.57 &  4.55 &	 $-$1.78& $-$1.82  \\
G 10-26         &  S *  &  5900  &  5915 &	 4.25 &  4.35 &	 $-$0.03& $-$0.13  \\
G 11-36         &  S *  &  5682  &  5583 &	 4.32 &  4.40 &	 $-$0.68& $-$0.56  \\
G 11-37         &  K *  &  5287  &  5228 &	 4.85 &  4.77 &	 $-$0.14& $-$0.19  \\
G 11-44         &  K    &  6010  &  6082 &	 4.18 &  4.18 &	 $-$2.07& $-$2.05  \\
G 11-45         &  K *  &  5490  &  5535 &	 4.82 &  4.93 &	 $-$0.01& $-$0.03  \\
G 13-1          &  S *  &  5817  &  5792 &	 4.67 &  4.58 &	 $-$0.25& $-$0.22  \\
G 13-9          &  K    &  6082  &  6270 &	 4.24 &  4.22 &	 $-$2.31& $-$2.26  \\
G 13-38         &  S *  &  5220  &  5204 &	 4.60 &  4.66 &	 $-$0.96& $-$0.96  \\
G 14-5          &  S *  &  5342  &  5507 &	 4.65 &  4.63 &	 $-$0.70& $-$0.57  \\
G 14-24         &  K    &  4970  &  4815 &	 4.76 &  4.85 &	 $-$2.17& $-$2.22  \\  
G 14-26         &  S    &  5800  &  5772 &	 4.40 &  4.66 &	 $-$0.20& $-$0.20  \\  
G 14-38         &  S    &  5235  &  5180 &	 4.75 &  4.76 &	 $-$0.42& $-$0.24  \\  
G 14-54         &  S    &  5612  &  5490 &	 4.57 &  4.73 &	 $-$0.13& $-$0.14  \\  
G 16-9          &  E *  &  4892  &  4903 &	 4.89 &  4.89 &	 $-$0.77& $-$0.72  \\  
G 16-13         &  S    &  5562  &  5460 &	 4.40 &  4.40 &	 $-$1.03& $-$1.01  \\  
G 16-31         &  S *  &  4890  &  4753 &	 4.82 &  4.84 &	 $-$0.55& $-$0.63  \\  
G 17-16         &  E    &  5190  &  5216 &	 4.61 &  4.50 &	 $-$0.83& $-$0.76  \\  
G 17-21         &  L *  &  5835  &  5873 &	 4.27 &  4.44 &	 $-$0.66& $-$0.64  \\  
G 17-29         &  E *  &  5322  &  5311 &	 4.95 &  5.23 &	 $+$0.05& $+$0.02  \\
G 17-30         &  S *  &  5600  &  5672 &	 4.57 &  4.69 &	 $-$0.48& $-$0.42  \\  
G 18-40         &  K    &  5800  &  5778 &	 4.24 &  4.46 &	 $-$1.76& $-$1.75  \\  
G 20-8          &  E    &  5957  &  6066 &	 4.23 &  4.19 &	 $-$2.30& $-$2.27  \\  
G 22-20         &  E    &  6060  &  5898 &	 3.86 &  4.32 &	 $-$0.82& $-$1.07  \\  
G 23-14         &  K    &  4922  &  5000 &	 4.87 &  4.60 &	 $-$2.05& $-$1.92  \\  
G 23-16         &  K *  &  4900  &  4773 &	 4.97 &  5.01 &	 $-$0.03& $-$0.02  \\  
G 24-15         &  L *  &  5912  &  5926 &	 4.09 &  4.06 &	 $-$1.10& $-$1.12  \\  
G 24-17         &  K *  &  4777  &  4639 &	 4.78 &  4.60 &	 $-$0.95& $-$0.94  \\  
G 24-18F        &  K *  &  4767  &  4602 &	 4.91 &  4.74 &	 $-$0.87& $-$0.84  \\  
G 28-31         &  L    &  5762  &  6009 &	 4.39 &  4.34 &	 $-$2.22& $-$2.16  \\
G 29-25         &  K    &  5340  &  5372 &	 4.46 &  4.54 &	 $-$0.99& $-$0.86  \\  
G 29-71         &  K    &  5685  &  5585 &	 4.48 &  4.26 &	 $-$2.26& $-$2.34  \\  
G 31-26         &  K    &  5345  &  5389 &	 4.51 &  4.84 &	 $-$1.49& $-$1.51  \\  
G 37-26         &  L *  &  5940  &  5860 &	 4.21 &  4.08 &	 $-$1.93& $-$1.87  \\  
G 40-14         &  S    &  6257  &  6260 &	 4.20 &  3.98 &	 $-$2.54& $-$2.55  \\  
G 43-5          &  K    &  5310  &  5210 &	 4.66 &  4.58 &	 $-$2.12& $-$2.09  \\  
G 43-30         &  K *  &  5120  &  5108 &	 5.05 &  4.92 &	 $-$0.12& $+$0.02  \\  
G 43-33         &  L *  &  5925  &  5899 &	 4.30 &  4.42 &	 $-$0.37& $-$0.26  \\  
G 43-44         &  K *  &  5307  &  5114 &	 4.90 &  5.07 &	 $-$0.08& $-$0.06  \\  
G 44-6          &  S *  &  5617  &  5629 &	 4.61 &  4.60 &	 $-$0.54& $-$0.63  \\  
G 44-30         &  S    &  5425  &  5405 &	 4.49 &  4.62 &	 $-$0.89& $-$0.88  \\  
G 46-31         &  S    &  5772  &  5706 &	 4.29 &  4.42 &	 $-$0.89& $-$0.90  \\  
G 48-29         &  L    &  6257  &  6373 &	 4.22 &  4.16 &	 $-$2.50& $-$2.53  \\  
G 53-30         &  S    &  5450  &  5492 &	 4.57 &  4.65 &	 $-$0.43& $-$0.39  \\  
G 53-41         &  L    &  5967  &  5951 &	 3.91 &  3.73 &	 $-$1.21& $-$1.20  \\  
G 54-21         &  K *  &  5862  &  5788 &	 4.48 &  4.64 &	 $-$0.03& $-$0.17  \\  
G 56-48         &  K    &  4775  &  4673 &	 4.78 &  4.85 &	 $-$2.20& $-$2.20  \\  
G 58-23         &  K *  &  5540  &  5456 &	 4.40 &  4.43 &	 $-$0.97& $-$0.90  \\  
G 58-25         &  K *  &  5930  &  5908 &	 4.01 &  4.07 &	 $-$1.41& $-$1.37  \\  
G 58-30         &  K *  &  5855  &  5803 &	 4.70 &  4.62 &	 $+$0.30:& $-$0.05:  \\  
G 58-41         &  K *  &  5865  &  5869 &	 4.39 &  4.33 &	 $-$0.33& $-$0.46  \\  
G 59-1          &  K    &  5430  &  5353 &	 4.56 &  4.43 &	 $-$1.02& $-$0.74  \\  
G 59-27         &  K    &  6092  &  6136 &	 4.09 &  4.16 &	 $-$2.10& $-$2.08  \\  
G 60-46         &  S    &  5300  &  5160 &	 4.59 &  4.59 &	 $-$1.19& $-$1.28  \\  
G 60-48         &  L    &  5817  &  5835 &	 4.28 &  4.15 &	 $-$1.63& $-$1.62  \\  
G 62-44         &  S *  &  5102  &  5024 &	 4.87 &  4.81 &	 $-$0.58& $-$0.29  \\  
G 62-52         &  K *  &  5252  &  5224 &	 4.59 &  4.62 &	 $-$1.28& $-$1.36  \\  
G 62-61         &  S *  &  5830  &  5733 &	 4.52 &  4.39 &	 $-$0.32& $-$0.31  \\  
G 64-12         &  K    &  6272  &  6354 &	 4.28 &  4.13 &	 $-$3.31& $-$3.16  \\  
G 64-37         &  E    &  6377  &  6367 &	 4.20 &  4.01 &	 $-$3.00& $-$3.04  \\  
G 64-54         &  S *  &  5332  &  5361 &	 4.89 &  4.91 &	 $-$0.10& $-$0.28  \\  
G 65-47         &  S *  &  5607  &  5598 &	 4.58 &  4.72 &	 $-$0.35& $-$0.36  \\  
G 66-9          &  E    &  5685  &  5747 &	 4.49 &  4.51 &	 $-$2.23& $-$2.24  \\  
G 66-15         &  S *  &  5590  &  5375 &	 4.53 &  4.61 &	 $-$0.20& $-$0.46  \\  
G 66-49         &  S *  &  5345  &  5478 &	 4.74 &  4.86 &	 $-$0.57:& $-$0.13:  \\  
G 75-56         &  K    &  6040  &  6238 &	 4.19 &  4.13 &	 $-$2.33& $-$2.21  \\  
G 79-42         &  K    &  5635  &  5655 &	 4.32 &  4.36 &	 $-$1.10& $-$1.16  \\
G 80-15         &  K *  &  5800  &  5823 &	 4.23 &  4.35 &	 $-$0.78& $-$0.81  \\  
G 82-42         &  K    &  5535  &  5453 &	 4.23 &  4.08 &	 $-$1.16& $-$1.14  \\  
G 82-47         &  E *  &  4837  &  4765 &	 4.92 &  5.09 &	 $-$0.45& $-$0.52  \\  
G 84-37         &  K *  &  5945  &  5896 &	 4.17 &  4.14 &	 $-$0.81& $-$0.80  \\  
G 89-14         &  K    &  5917  &  5962 &	 4.19 &  4.09 &	 $-$1.76& $-$1.57  \\  
G 90-25         &  K *  &  5392  &  5296 &	 4.52 &  4.73 &	 $-$1.62& $-$1.67  \\  
G 92-15         &  K *  &  5725  &  5684 &	 4.55 &  4.57 &	 $-$0.11& $-$0.25  \\ 
G 92-6          &  S    &  6127  &  6192 &	 4.23 &  4.15 &	 $-$2.68& $-$2.66  \\  
G 99-40         &  K    &  5970  &  5856 &	 4.08 &  4.30 &	 $-$0.35& $-$0.41  \\  
G 99-48         &  K    &  5077  &  5044 &	 4.70 &  4.46 &	 $-$1.92& $-$1.96  \\  
G 106-46        &  S    &  5842  &  5864 &	 4.29 &  4.47 &	 $-$0.51& $-$0.49  \\  
G 108-33        &  K    &  6082  &  6222 &	 4.33 &  4.27 &	 $-$2.69& $-$2.81  \\  
G 108-53        &  S *  &  5645  &  5583 &	 4.45 &  4.40 &	 $-$0.57& $-$0.56  \\  
G 110-34        &  K    &  6105  &  6146 &	 3.83 &  4.14 &	 $-$1.58& $-$1.54  \\  
G 112-1         &  S    &  5425  &  5213 &	 4.55 &  4.77 &	 $-$2.57& $-$2.58  \\  
G 113-22        &  S    &  5525  &  5522 &	 4.39 &  4.47 &	 $-$1.21& $-$1.26  \\  
G 114-18        &  S *  &  5545  &  5528 &	 4.73 &  4.64 &	 $+$0.05& $-$0.16  \\  
G 114-26        &  S *  &  5837  &  5844 &	 4.31 &  4.24 &	 $-$1.78& $-$1.72  \\  
G 114-48        &  S    &  5555  &  5544 &	 4.52 &  4.67 &	 $-$0.41& $-$0.27  \\  
G 121-12        &  K    &  5955  &  5964 &	 4.03 &  4.09 &	 $-$0.92& $-$0.96  \\  
G 125-64        &  K *  &  5860  &  5697 &	 4.28 &  4.27 &	 $-$1.92& $-$2.11  \\  
G 126-36        &  K    &  5555  &  5582 &	 4.54 &  4.50 &	 $-$0.91& $-$0.91  \\  
G 126-52        &  E    &  6302  &  6302 &	 4.05 &  4.06 &	 $-$2.41& $-$2.47  \\  
G 137-87        &  S    &  5755  &  5890 &	 4.45 &  4.64 &	 $-$2.62& $-$2.68  \\  
G 139-49        &  E    &  5315  &  5336 &	 4.59 &  4.51 &	 $-$1.11& $-$0.99  \\  
G 141-19        &  E    &  5135  &  5212 &	 4.85 &  4.54 &	 $-$2.43& $-$2.42  \\  
G 143-27        &  K    &  5670  &  5708 &	 4.37 &  4.23 &	 $-$1.62& $-$1.61  \\  
G 151-59        &  S *  &  5167  &  4941 &	 4.95 &  5.01 &	 $+$0.03& $-$0.05  \\  
G 152-67        &  K    &  5227  &  5180 &	 4.72 &  4.40 &	 $-$2.47& $-$2.53  \\  
G 160-3         &  E *  &  5575  &  5657 &	 4.76 &  4.89 &	 $-$0.14& $-$0.14  \\  
G 161-14        &  S    &  5652  &  5690 &	 4.26 &  4.53 &	 $-$1.10& $-$0.89  \\  
G 161-73        &  S    &  5797  &  5821 &	 4.23 &  4.26 &	 $-$1.29& $-$1.10  \\  
G 162-16        &  S    &  5690  &  5658 &	 4.34 &  4.56 &	 $-$0.53& $-$0.33  \\  
G 162-51        &  S    &  5765  &  5724 &	 4.16 &  4.48 &	 $-$0.52& $-$0.62  \\  
G 162-68        &  S *  &  5385  &  5177 &	 4.67 &  4.77 &	 $-$0.54& $-$0.72  \\  
G 163-70        &  S    &  5805  &  5768 &	 4.19 &  4.41 &	 $-$1.25& $-$1.23  \\  
G 165-11        &  K *  &  5785  &  5817 &	 4.32 &  4.38 &	 $-$0.46& $-$0.66  \\  
G 166-45        &  E *  &  5997  &  6187 &	 4.30 &  4.36 &	 $-$2.35& $-$2.30  \\  
G 170-47        &  E    &  5225  &  5191 &	 4.82 &  4.62 &	 $-$2.59& $-$2.58  \\  
G 171-50        &  K    &  5320  &  5100 &	 4.70 &  4.60 &	 $-$1.97& $-$1.94  \\  
G 180-58        &  K *  &  5090  &  5013 &	 4.76 &  4.71 &	 $-$2.14& $-$2.15  \\  
G 186-26        &  E    &  6215  &  6336 &	 4.22 &  4.23 &	 $-$2.64& $-$2.74  \\  
G 195-52        &  K *  &  5332  &  5384 &	 4.90 &  4.87 &	 $-$0.10& $-$0.10  \\  
G 196-48        &  K    &  5690  &  5634 &	 4.30 &  3.73 &	 $-$1.74& $-$1.50  \\  
G 206-34        &  E    &  6170  &  6241 &	 4.27 &  4.22 &	 $-$2.62& $-$2.74  \\  
G 209-35        &  K *  &  5070  &  5069 &	 4.88 &  4.85 &	 $-$0.49& $-$0.33  \\  
G 229-34        &  K *  &  5527  &  5576 &	 4.60 &  4.59 &	 $-$0.50& $-$0.27  \\
G 236-11        &  K *  &  5970  &  5821 &	 4.51 &  4.56 &	 $+$0.31:& $-$0.10:  \\  
G 271-34        &  L *  &  5647  &  5669 &	 4.48 &  4.56 &	 $-$0.68& $-$0.76  \\  
HD 693          &  E *  &  6120  &  6010 &	 4.08 &  4.36 &	 $-$0.38& $-$0.53  \\  
HD 3567         &  E    &  5990  &  6022 &	 4.50 &  4.38 &	 $-$1.29& $-$1.27  \\  
HD 4306         &  L    &  4815  &  4701 &	 2.40 &  2.28 &	 $-$2.71& $-$2.80  \\  
HD 6268         &  L    &  4695  &  4731 &	 2.07 &  2.01 &	 $-$2.48& $-$2.55  \\  
HD 6461         &  L    &  4810  &  5147 &	 2.68 &  2.80 &	 $-$0.93& $-$0.87  \\  
HD 6833         &  K    &  4707  &  4682 &	 2.54 &  2.67 &	 $-$0.93& $-$0.96  \\  
HD 8724         &  K    &  4680  &  4760 &	 2.00 &  1.96 &	 $-$1.64& $-$1.69  \\  
HD 16031        &  E    &  6005  &  6009 &	 4.12 &  4.22 &	 $-$1.71& $-$1.73  \\  
HD 59392        &  L    &  5892  &  5905 &	 4.22 &  4.05 &	 $-$1.65& $-$1.63  \\  
HD 74000        &  L    &  6075  &  6056 &	 4.14 &  3.79 &	 $-$1.82& $-$1.87  \\  
HD 76932        &  L *  &  5860  &  5840 &	 4.02 &  4.17 &	 $-$0.99& $-$0.91  \\  
HD 83212        &  K    &  4575  &  4738 &	 1.98 &  2.08 &	 $-$1.48& $-$1.45  \\  
HD 84937        &  K *  &  6180  &  6293 &	 4.09 &  4.09 &	 $-$2.06:& $-$2.37:  \\  
HD 85773        &  K    &  4470  &  4654 &	 0.99 &  0.92 &	 $-$2.27& $-$2.23  \\  
HD 87140        &  K    &  4822  &  4909 &	 2.79 &  2.99 &	 $-$1.71& $-$1.81  \\  
HD 89499        &  S    &  4780  &  4737 &	 2.39 &  2.39 &	 $-$2.15& $-$2.21  \\  
HD 92588        &  K *  &  4942  &  5010 &	 4.65 &  5.03 &	 $-$0.07& $-$0.12  \\  
HD 93529        &  K    &  4810  &  4777 &	 2.32 &  2.34 &	 $-$1.67& $-$1.71  \\  
HD 97320        &  L *  &  5935  &  5872 &	 4.13 &  4.08 &	 $-$1.18& $-$1.15  \\  
HD 97916        &  L    &  6132  &  6337 &	 3.73 &  3.63 &	 $-$1.20& $-$1.12  \\  
HD 101063       &  L    &  4865  &  4991 &	 2.95 &  3.18 &	 $-$1.15& $-$1.13  \\  
HD 102644       &  K    &  6157  &  6052 &	 4.44 &  4.39 &	 $-$1.83& $-$1.86  \\  
HD 103545       &  K    &  4835  &  4837 &	 2.48 &  2.53 &	 $-$2.14& $-$2.17  \\  
HD 105546       &  K    &  4727: &  5095:&	 2.49 &  2.53 &	 $-$1.40& $-$1.33  \\  
HD 107752       &  K    &  4710  &  4787 &	 2.07 &  2.14 &	 $-$2.74& $-$2.69  \\  
HD 108317       &  K    &  5310  &  5179 &	 3.33 &  3.39 &	 $-$2.27& $-$2.30  \\  
HD 108405       &  S *  &  5705  &  5676 &	 4.48 &  4.48 &	 $-$0.60& $-$0.87  \\  
HD 110184       &  L    &  4360  &  4582 &	 0.80 &  0.84 &	 $-$2.46& $-$2.38  \\  
HD 113083       &  E *  &  5737  &  5605 &	 4.20 &  4.54 &	 $-$1.04& $-$1.09  \\  
HD 115444       &  K    &  4757  &  4830 &	 2.16 &  1.93 &	 $-$2.73& $-$2.72  \\  
HD 115772       &  L    &  4930  &  5133 &	 2.56 &  2.82 &	 $-$0.70& $-$0.70  \\  
HD 116064       &  S *  &  5862  &  5957 &	 4.37 &  4.26 &	 $-$1.91& $-$2.01  \\  
HD 117220       &  L    &  4895  &  5238 &	 2.68 &  2.53 &	 $-$0.86& $-$0.85  \\  
HD 122196       &  L    &  5905  &  5913 &	 4.30 &  4.15 &	 $-$1.89& $-$1.87  \\  
HD 122563       &  L    &  4687  &  4746 &	 1.61 &  1.54 &	 $-$2.62& $-$2.57  \\  
HD 122956       &  L    &  4600  &  4630 &	 1.81 &  1.93 &	 $-$1.75& $-$1.74  \\  
HD 126778       &  K    &  4807  &  4897 &	 2.60 &  2.63 &	 $-$0.59& $-$0.44  \\  
HD 128188       &  K    &  4677  &  4752 &	 2.04 &  2.13 &	 $-$1.37& $-$1.38  \\  
HD 132475       &  L *  &  5550  &  5564 &	 3.76 &  3.80 &	 $-$1.70& $-$1.62  \\  
HD 134169       &  L *  &  5782  &  5844 &	 4.26 &  4.24 &	 $-$0.85& $-$0.77  \\  
HD 134439       &  E    &  4950  &  4862 &	 4.66 &  4.52 &	 $-$1.53& $-$1.54  \\  
HD 134440       &  K    &  4732  &  4675 &	 4.73 &  4.52 &	 $-$1.37& $-$1.42  \\  
HD 136202       &  E *  &  6300  &  6088 &	 5.70:&  4.64:&	 $-$0.13& $-$0.16  \\  
HD 140283       &  L *  &  5792  &  5875 &	 3.75 &  3.70 &	 $-$2.47& $-$2.58  \\  
HD 142948       &  L    &  4647  &  4697 &	 1.86 &  2.20 &	 $-$0.89& $-$0.89  \\
HD 154417       &  E *  &  5880  &  6043 &	 4.43 &  4.66 &	 $-$0.18& $-$0.30  \\  
HD 161770       &  K *  &  5182  &  5335 &	 4.78 &  4.45 &	 $-$2.12& $-$2.01  \\  
HD 163810       &  E *  &  5570  &  5421 &	 4.32 &  4.56 &	 $-$1.34& $-$1.34  \\  
HD 166161       &  L    &  5125  &  5154 &	 1.84 &  1.89 &	 $-$1.22& $-$1.23  \\  
HD 184499       &  K *  &  5710  &  5801 &	 4.38 &  4.41 &	 $-$0.58& $-$0.46  \\  
HD 193901       &  E *  &  5655  &  5620 &	 4.40 &  4.50 &	 $-$1.08& $-$1.09  \\ 
HD 200654       &  E    &  5105  &  5065 &	 2.84 &  2.86 &	 $-$2.93& $-$2.91  \\
HD 201889       &  O *  &  5657  &  5621 &	 4.24 &  4.50 &	 $-$0.92& $-$0.82  \\  
HD 201891       &  L *  &  5830  &  5799 &	 4.20 &  4.30 &	 $-$1.13& $-$1.13  \\  
HD 210295       &  K    &  4725  &  4783 &	 2.48 &  2.71 &	 $-$1.36& $-$1.36  \\  
HD 211744       &  L    &  4865  &  4943 &	 3.03 &  3.15 &	 $-$1.03& $-$0.99  \\  
HD 216143       &  K    &  4622  &  4719 &	 1.51 &  1.45 &	 $-$2.16& $-$2.12  \\  
HD 218502       &  K *  &  5750  &  5673 &	 3.72 &  3.63 &	 $-$1.88& $-$1.99  \\  
HD 218857       &  L    &  5165: &  4740:&	 2.51 &  2.53 &	 $-$1.94& $-$1.91  \\  
HD 221170       &  K    &  4610  &  4686 &	 1.57 &  1.54 &	 $-$2.12& $-$2.11  \\  
LP 635-14       &  E    &  6045  &  6258 &	 4.31 &  4.26 &	 $-$2.80& $-$2.80  \\  
LP 685-44       &  E    &  5290: &  4726:&	 4.69 &  4.44 &	 $-$2.67& $-$2.62  \\  
LP 732-48       &  K    &  6122  &  6324 &	 4.25 &  4.18 &	 $-$2.46& $-$2.48  \\  
LP 831-70       &  E    &  6192  &  6312 &	 4.33 &  4.26 &	 $-$3.40& $-$3.25  \\  
LTT 2437        &  S    &  5677  &  5534 &	 4.52 &  4.76 &	 $-$2.56& $-$2.39  \\  
LTT 6194        &  S    &  5877  &  6029 &	 4.43 &  4.30 &	 $-$2.79& $-$2.86  \\ 
\enddata
\tablecomments{* indicates that the star is a member of the nearby
subsample \\ : indicates a large discrepancy between the catalog (CAT) and
network (ANN) parameter estimates; see appendix}
\end{deluxetable}


\begin{deluxetable}{llcccccc}
\tablenum{3}
\tablecolumns{8}
\tablewidth{0pt}
\tablecaption{Catalog and ANN Parameters for the Testing Sample
\label{tab-testlist}}
\tablehead{
\colhead{Star}                    & 
\colhead{Source}                  &
\colhead{T$_{\rm eff CAT}$} & 
\colhead{T$_{\rm eff ANN}$} & 
\colhead{$\log g_{\rm CAT}$}      &  
\colhead{$\log g_{\rm ANN}$}      & 
\colhead{[Fe/H]$_{\rm CAT}$}      & 
\colhead{[Fe/H]$_{\rm ANN}$}      \\
\colhead{}                        & 
\colhead{}                        &
\colhead{(K)}                     & 
\colhead{(K)}                     & 
\colhead{(dex)}                   & 
\colhead{(dex)}                   & 
\colhead{(dex)}                   & 
\colhead{(dex)}   }
\startdata 
BD +01 2916      &  L      & 4247: &  4782:&	 1.02:&  1.83:&	 $-$1.82:& $-$2.37: \\  
BD +29 2091      &  K *    & 5740  &  5660 &	 4.36 &  4.39 &	 $-$1.98& $-$1.77 \\  
BD $-$04 680     &  K      & 5650: &  5902:&	 4.53 &  4.20 &	 $-$2.22:& $-$1.81: \\  
BD $-$09 5746    &  E      & 5960  &  5942 &	 4.15 &  4.20 &	 $-$1.73& $-$1.85 \\  
BD $-$14 5890    &  K      & 4767  &  4925 &	 2.27:&  3.01:&	 $-$2.07& $-$2.05 \\  
CS 22873-128    &  E      & 4882  &  4779 &	 2.50:&  3.37:&	 $-$2.88& $-$2.98 \\  
CS 22891-200    &  L      & 4632: &  5053:&	 1.87:&  4.02: & $-$3.49:& $-$2.88: \\  
CS 22949-048    &  P      & 4665  &  4858 &	 1.95 &  2.13 &	 $-$3.17& $-$2.98 \\  
CS 22968-014    &  L      & 4815: &  5335:&	 2.24 &  2.96 &	 $-$3.43:& $-$2.94: \\  
G 13-35         &  L *    & 6055  &  6145 &	 4.08 &  3.85 &	 $-$1.63& $-$1.82 \\  
G 14-41         &  S      & 5350  &  5410 &	 4.74 &  4.82 &	 $-$0.34& $-$0.16 \\  
G 15-6          &  S *    & 5295  &  5265 &	 4.65 &  4.61 &	 $-$0.65& $-$0.63 \\  
G 15-14         &  S      & 5102  &  4925 &	 4.77 &  4.89 &	 $-$0.37& $-$0.37 \\  
G 15-17         &  S *    & 5067  &  4954 &	 4.88 &  4.91 &	 $-$0.39& $-$0.31 \\  
G 17-22         &  S *    & 4765: &  5687:&	 4.90 &  4.42 &	 $-$0.77& $-$0.80 \\  
G 20-24         &  E      & 6052  &  5974 &	 4.12 &  4.23 &	 $-$2.07& $-$2.29 \\  
G 21-22         &  E      & 6167: &  5828:&	 3.70:&  4.64:&	 $-$0.88:& $-$1.18: \\  
G 28-42         &  K      & 5397  &  5143 &	 4.46 &  4.59 &	 $-$1.57& $-$1.58 \\  
G 44-44         &  K *    & 5637  &  5726 &	 4.66 &  4.57 &	 $-$0.16& $-$0.13 \\  
G 54-7          &  K *    & 5887  &  5862 &	 4.34 &  4.52 &	 $-$0.16& $-$0.23 \\  
G 56-30         &  S      & 5842  &  5656 &	 4.19 &  4.45 &	 $-$0.91& $-$1.00 \\  
G 57-11         &  K *    & 5570  &  5664 &	 4.84 &  4.71 &	 $+$0.03& $-$0.08 \\  
G 59-24         &  S      & 5995  &  5828 &	 4.26 &  4.07 &	 $-$2.42& $-$2.50 \\  
G 60-66         &  S *    & 5437  &  5537 &	 4.73 &  4.76 &	 $-$0.26& $-$0.28 \\  
G 63-46         &  K      & 5625  &  5744 &	 4.39 &  4.30 &	 $-$0.91& $-$0.75 \\  
G 66-65         &  K *    & 5727  &  5596 &	 4.21 &  4.32 &	 $-$0.78& $-$0.78 \\  
G 84-29         &  E      & 6355  &  6326 &	 4.14 &  4.20 &	 $-$2.67& $-$2.79 \\  
G 84-39         &  E *    & 5055  &  5086 &	 4.83 &  5.01 &	 $-$0.66& $-$0.61 \\  
G 90-3          &  K      & 5842  &  5821 &	 3.86 &  3.99 &	 $-$2.18& $-$2.22 \\  
G 97-43         &  E      & 5215  &  5216 &	 4.71 &  4.77 &	 $-$0.49& $-$0.31 \\  
G 99-52         &  S      & 5082  &  5172 &	 4.64 &  4.38 &	 $-$1.40:& $-$2.01: \\  
G 106-53        &  S *    & 4955  &  5135 &	 4.88 &  4.81 &	 $-$0.21:& $-$0.58: \\  
G 113-24        &  S *    & 5737  &  5711 &	 4.34 &  4.46 &	 $-$0.49& $-$0.59 \\  
G 114-19        &  S *    & 5265  &  5553 &	 4.80 &  4.42 &	 $-$0.42& $-$0.58 \\  
G 122-43        &  K      & 5570  &  5689 &	 4.58 &  4.34 &	 $-$2.36& $-$2.25 \\  
G 139-8         &  E      & 5997  &  5894 &	 4.31 &  4.34 &	 $-$2.36& $-$2.60 \\  
G 141-15        &  S      & 5955  &  6181 &	 4.32 &  4.50 &	 $-$2.67& $-$2.57 \\  
G 146-76        &  K      & 5150  &  5046 &	 4.69: &  3.57: & $-$2.15& $-$2.06 \\  
G 154-32        &  E *    & 5765  &  5822 &	 4.54 &  4.58 &	 $-$0.19& $-$0.32 \\  
G 161-84        &  S      & 4605: &  5013:&	 4.72 &  4.60 &	 $-$1.57& $-$1.32 \\  
G 200-62        &  K *    & 5080  &  5086 &	 4.84 &  4.86 &	 $-$0.45& $-$0.44 \\  
Groom 1830      &  O      & 5010  &  4846 &	 4.63 &  4.54 &	 $-$1.31& $-$1.47 \\  
HD 3008         &  K      & 4370  &  4720 &	 0.99 &  0.50 &	 $-$1.90& $-$1.87 \\  
HD 6755         &  K      & 5230: &  4864:&	 2.98 &  3.55 &	 $-$1.49:& $-$1.98: \\  
HD 13979        &  L      & 4925  &  5072 &	 2.58 &  2.01 &	 $-$2.61& $-$2.71 \\  
HD 20010        &  E *    & 6077  &  6020 &	 4.72 &  4.38 &	 $-$0.27& $-$0.41 \\  
HD 20038        &  L      & 4875  &  4979 &	 2.41:&  3.21:&	 $-$0.87& $-$1.14 \\
HD 22484        &  K *    & 6080  &  5880 &	 5.02 &  4.57 &	 $-$0.16& $-$0.22 \\  
HD 34328        &  E *    & 5857  &  5809 &	 4.15 &  4.47 &	 $-$1.61& $-$1.59 \\  
HD 44007        &  L      & 4750  &  4733 &	 2.71:&  1.61:&	 $-$1.58& $-$1.85 \\  
HD 45282        &  K      & 4980  &  4990 &	 3.53 &  3.96 &	 $-$1.52:& $-$1.84: \\  
HD 63791        &  K      & 4762  &  4760 &	 2.21 &  2.00 &	 $-$1.67& $-$2.00 \\  
HD 74462        &  K      & 4812  &  4777 &	 2.91: &  1.89: & $-$1.42& $-$1.61 \\  
HD 99383        &  L *    & 5892  &  6085 &	 4.13 &  4.04 &	 $-$1.65& $-$1.79 \\  
HD 111721       &  L      & 4750  &  4511 &	 3.01:&  1.46:&	 $-$1.26:& $-$2.72: \\  
HD 111980       &  L *    & 5747  &  5679 &	 4.26 &  4.43 &	 $-$0.99& $-$1.15 \\  
HD 114762       &  K *    & 5860  &  5832 &	 4.23 &  4.49 &	 $-$0.70& $-$0.72 \\  
HD 128279       &  L      & 5130  &  5093 &	 3.11: &  4.54: & $-$2.20& $-$2.15 \\  
HD 149414       &  L *    & 5040  &  5001 &	 4.64 &  4.61 &	 $-$1.30& $-$1.29 \\
HD 160617       &  E      & 5955  &  5817 &	 4.15 &  4.12 &	 $-$1.78& $-$1.88 \\  
HD 181743       &  L *    & 5915  &  6076 &	 4.16 &  4.05 &	 $-$1.79& $-$2.02 \\  
HD 186478       &  K      & 4712  &  4776 &	 1.71 &  1.51 &	 $-$2.58& $-$2.49 \\  
HD 187111       &  K      & 4247: &  4688:&	 0.97 &  1.17 &	 $-$1.78& $-$1.79 \\  
HD 188510       &  K *    & 5470  &  5388 &	 4.39 &  4.51 &	 $-$1.53& $-$1.47 \\  
HD 195636       &  K      & 5487: &  5820:&	 3.27 &  3.75 &	 $-$2.80& $-$2.58 \\  
HD 196944       &  K      & 5122  &  5045 &	 2.89:&  1.57:&	 $-$2.33:& $-$1.95: \\  
HD 213657       &  E      & 6060  &  5986 &	 4.10 &  4.13 &	 $-$1.98& $-$2.13 \\  
HD 219617       &  E *    & 5907  &  5785 &	 4.17 &  4.29 &	 $-$1.31& $-$1.65 \\  
LP 815-43       &  E      & 6305  &  6299 &	 4.24 &  4.20 &	 $-$3.20:& $-$2.79: \\  
Ross 740        &         & 5010: &  5968:&	 4.75 &  4.39 &	 $-$2.75& $-$2.66 \\  
\enddata
\tablecomments{* indicates that the star is a member of the nearby
subsample \\ : indicates a large discrepancy between the catalog (CAT) and
network (ANN) parameter estimates; see appendix}
\end{deluxetable}


\begin{deluxetable}{lrrrrrr}
\label{tab-stats}
\tablenum{4}
\tablewidth{5.7in}
\tablecaption{Statistics of the ANN Results}
\tablecolumns{7}
\tablehead{
\colhead{Subsample}                        &
\colhead{$C_{BI}$}                         &
\colhead{$C_{BI}$}                         &
\colhead{$S_{BI}$}                         &
\colhead{$S_{BI}$}                         &
\colhead{Number}                              &
\colhead{Number}                             \\
\colhead{}                                 &
\colhead{training}                              &
\colhead{testing}                              &
\colhead{training}                              &
\colhead{testing}                              &
\colhead{training}                             &
\colhead{testing}                              
}
\startdata
\cutinhead{T$_{\rm eff}$}

nearby/kpno&    --4&    --38&       37&       139&     30&     10 \\
nearby/full&   --12&     +39&       67&       219&     76&     25 \\
total/kpno &     +4&     +77&       72&       215&     86&     28 \\
total/full &     +7&      +3&      110&       185&    209&     70 \\

\cutinhead{log g}

nearby/kpno& --0.03&   +0.00&    0.10&       0.17&     30&     10 \\
nearby/full&   0.00&   +0.02&    0.07&       0.17&     76&     25 \\
total/kpno &  +0.02&   +0.04&    0.16&       0.32&     86&     28 \\ 
total/full &  +0.01&    0.00&    0.15&       0.36&    209&     70 \\

\cutinhead{[Fe/H]}

nearby/kpno&   0.00&  --0.16&    0.08&       0.16&     30&     10 \\
nearby/full& --0.01&   +0.04&    0.05&       0.24&     76&     25 \\
total/kpno & --0.01&   +0.00&    0.13&       0.30&     86&     28 \\
total/full &   0.00&  --0.05&    0.09&       0.21&    209&     70 \\

\enddata
\end{deluxetable}


\begin{deluxetable}{lrrrrrr}
\label{tab-noisystats}
\tablenum{5}
\tablewidth{5.7in}
\tablecaption{Statistics of the ANN Results -- S/N Experiments}
\tablecolumns{7}
\tablehead{
\colhead{Subsample}                        &
\colhead{$C_{BI}$}                         &
\colhead{$C_{BI}$}                         &
\colhead{$S_{BI}$}                         &
\colhead{$S_{BI}$}                         &
\colhead{Number}                              &
\colhead{Number}                             \\
\colhead{}                                 &
\colhead{training}                              &
\colhead{testing}                              &
\colhead{training}                              &
\colhead{testing}                              &
\colhead{training}                             &
\colhead{testing}                              
}
\startdata
\cutinhead{Trained on High S/N Spectra}

\cutinhead{T$_{\rm eff}$}

S/N$>$40 &       +59 &    --16 &     116  &    144   &     37  &   15 \\
S/N = 26 &       +42 &    --47 &     123  &    142   &     37  &   15 \\
S/N = 13 &       +46 &    --42 &     115  &    165   &     37  &   15 \\

\cutinhead{log g}

S/N$>$40 &    --0.04 &   +0.04 &   0.12   &    0.23  &     37  &   15 \\
S/N = 26 &    --0.25 &  --0.20 &   0.19   &    0.29  &     37  &   15 \\
S/N = 13 &    --0.61 &  --0.66 &   0.36   &    0.50  &     37  &   15 \\    

\cutinhead{[Fe/H]}

S/N$>$40 &      0.00  &  +0.03 &    0.08  &    0.24  &     37  &   15 \\
S/N = 26 &     --0.38 & --0.41 &    0.12  &    0.26  &     37  &   15 \\
S/N = 13 &     --0.75 & --0.80 &    0.19  &    0.33  &     37  &   15 \\

\cutinhead{Trained on Similar S/N Spectra}

\cutinhead{T$_{\rm eff}$}

S/N$>$40 &       +2  &   +32  &     38   &    121  &    38  &   14 \\
S/N = 26 &      --8  &   +37  &     38   &    184  &    38  &   14 \\
S/N = 13 &      --1  &   +18  &     40   &    131  &    38  &   14 \\

\cutinhead{log g}

S/N$>$40 &     +0.01 &  +0.10 &    0.09  &    0.39  &   38  &   14 \\
S/N = 26 &      0.00 & --0.02 &    0.10  &    0.40  &   38  &   14 \\
S/N = 13 &      0.00 &  +0.05 &    0.10  &    0.31  &   38  &   14 \\ 

\cutinhead{[Fe/H]}

S/N$>$40 &    --0.02 &  +0.11 &   0.06   &    0.22  &   38  &   14 \\
S/N = 26 &     +0.01 &  +0.13 &   0.07   &    0.31  &   38  &   14 \\
S/N = 13 &    --0.01 &  +0.07 &   0.05   &    0.21  &   38  &   14 \\

\enddata
\end{deluxetable}

\clearpage

\end{document}